\documentclass{aa}

\usepackage{graphicx}
\usepackage{epsfig}
\usepackage{lscape}
\usepackage{rotating}
\usepackage{natbib}
\usepackage{tabularx}
\usepackage{subfig}
\bibpunct{(}{)}{;}{a}{}{,}
\usepackage{txfonts}
\begin{document}

   \title{The Infrared Massive Stellar Content of M83
     \thanks{This paper includes data gathered with the 6.5 meter Magellan 
       Telescopes located at Las Campanas Observatory, Chile.},
     \thanks{Based on observations with the NASA/ESA \textit{Hubble Space 
       Telescope}, obtained at the Space Telescope Science Institute,
       which is operated by the Association of Universities for Research
       in Astronomy, Inc., under NASA contract NAS5-26555.}}


   \author{S. J. Williams
          \inst{1}
          \and
	  A. Z. Bonanos
	  \inst{1}
	  \and
	  B. C. Whitmore
	  \inst{2}
	  \and
	  J. L. Prieto
	  \inst{3,4}
	  \and
	  W. P. Blair
	  \inst{5}}

   \institute{IAASARS, National Observatory of Athens, GR-15236 Penteli, Greece
             \email{williams@noa.gr}
         \and
             Space Telescope Science Institute, 3700 San Martin Drive,
	     Baltimore, MD 21218, USA
	 \and
	     N\'ucleo de Astronom\'ia de la Facultad de Ingenier\'ia,
	     Universidad Diego Portales, Av. Ej\'ercito 441, Santiago, Chile
	 \and
	     Millennium Institute of Astrophysics, Santiago, Chile
	 \and
	     The Henry A. Rowland Department of Physics and Astronomy, 
	     Johns Hopkins University, 3400 N. Charles Street, Baltimore, 
	     MD 21218, USA}

   \date{Received ; accepted }

\authorrunning{Williams et al.}
\titlerunning{Massive Stellar Content of M83}

\abstract
{}
{We present an analysis of archival \textit{Spitzer} images and new 
ground-based and \textit{Hubble Space Telescope} (\textit{HST}) 
near-infrared (IR) 
and optical images of the field of M83 with the goal of
identifying rare, dusty, evolved massive stars.}
{We present point source catalogs 
consisting of 3778 objects from \textit{Spitzer} Infrared Array Camera (IRAC) 
Band 1 (3.6 $\mu$m) and Band 2 (4.5 $\mu$m), and 975 objects identified 
in Magellan 6.5m FourStar near-IR $J$ and $K_{\rm s}$ images. A combined
catalog of coordinate matched near- and mid-IR point sources yields 221
objects in the field of M83.}
{We find 49 strong candidates for massive stars which
are very promising objects for spectroscopic follow-up. Based
on their location in a $B-V$ versus $V-I$ diagram, we expect at least 
24, or roughly 50\%, to be confirmed as red supergiants.}
{}


\maketitle

%

\section{Introduction}

Massive stars are of prime importance in astrophysics. The consequences
of their birth and evolution have profound ramifications for galaxy
evolution. Throughout their lives, their physical characteristics
shape and mold their immediate environment via intense radiation 
fields and strong winds. Many massive stars expire as supernovae (SNe), 
thus affecting subsequent star formation
and the chemical enrichment of galaxies, their interstellar media,
and the intergalactic medium.
Predicting the evolution of a massive star relies heavily upon the
mass loss experienced by a star over the course of its life. 

Mass loss in massive stars ($M\gtrsim8M_{\odot}$), particularly in
the late stages of evolution, is poorly understood 
\citep[see the recent review by][]{smi14}. Wind mass loss rates of
massive stars on the main sequence, once
thought to be a major contributor to massive star mass loss, are
now being revised to be 2 to 3 times lower than previous estimates 
based on observations in the ultraviolet and optical 
\citep[see][for example]{cro02,rep04,pul06} as well as in the X-ray 
\citep{coh11}. 
Episodic and eruptive mass loss in evolved massive stars must
therefore be more important than previously thought. 

Understanding mass loss via observations of Galactic massive stars is 
difficult. Massive stars are rare and are formed in regions 
along the Galactic plane where extinction can be patchy and very high. 
For example, \citet{ram12} estimate visual
extinctions up to 21.5 mag for members of the young massive stellar cluster
Masgomas-1 at a distance of $\sim$3.5 kpc. In order to observe massive
stars at large distances in the Milky Way, it becomes necessary to observe
them in the infrared (IR). By far the best efforts to discover
new Galactic massive stars exhibiting heavy mass loss 
have been most successful using \textit{Spitzer} 24 $\mu$m observations 
that revealed circumstellar shells of ejected material surrounding 
massive evolved stars \citep[e.g.][]{gav10,wac10}. In order to form
a more complete census, however, we must look outside our own Galaxy.

The consequences of late evolutionary stage mass loss for lower mass
stars may be seen in the progenitors of some supernovae. 
One specific example is SN 2012aw, a type IIP event in the galaxy 
M95 \citep{fag12}. Pre-explosion mass loss is likely 
responsible for the dusty circumstellar material (CSM) seen as a
significant visual extinction around the
red supergiant (RSG) progenitor that has a mass at the high end
of the range for type IIP SN progenitors \citep{fra12,van12}.

The situation concerning mass loss in more massive stars is uncertain. 
SN 2009ip was initially mistakenly identified 
as a SN \citep{maz09} based on its initial rise in brightness. 
However, it never reached typical 
SNe magnitudes, and was eventually understood to
be an object undergoing a luminous blue variable (LBV) eruption 
\citep{ber09,mil09}. SN 2009ip also experienced a similar outburst in
2010, followed by a brighter event in 2012. The jury is still
out concerning whether SN 2009ip is a true SN 
\citep{mau13,smi14,mau14,gra14} or an imposter \citep{pas13,mar14,fra15},
but poorly understood mass loss events have clearly played a vital 
role in shaping the late stages of evolution for this $50-80 M_{\odot}$ star.
The specifics of episodic and
eruptive mass loss mechanisms in both the higher mass 
\citep[$\gtrsim30M_{\odot}$,][]{hum94,hum14} and lower mass regimes
\citep[RSGs with masses $\lesssim30M_{\odot}$,][]{smi14} remain unknown.
While the case of SN 2009ip is rare, discovering and cataloging 
all massive stars with mass loss is a tractable investigation.

Mass loss in massive stars can reveal itself via an IR 
excess in different ways depending on the type and mass of the star. In a
recent study of luminous and variable stars in M31 and M33, \citet{hum14}
found that LBVs had no hot or warm dust evident in the IR. \ion{Fe}{ii}
emission stars have an IR excess attributed to circumstellar dust 
and nebulosity, both of which are indicative of mass loss. Meanwhile, 
only some A-F spectral type supergiants show IR evidence of mass loss, 
possibly because some are post-RSG objects, while others are still 
evolving to the red on the Hertzsprung-Russel (HR) diagram. For redder 
objects (RSGs), mass loss is seen as thermal-IR excess from hot dust
and molecular emission \citep{smi14}, and masers in the extreme cases
\citep{hab96}.

Work categorizing
the massive stellar content of nearby galaxies in the IR with 
\textit{Spitzer} began with the Large
Magellanic Cloud \citep[LMC;][]{blu06,bon09}, Small Magellanic Cloud 
\citep[SMC;][]{bon10} and M33 \citep{tho09}. \citet{bon09,bon10}
laid the foundation for differentiating the types of massive
stars by looking at objects with previously known spectral 
classifications in the LMC and SMC.
\citet{tho09} focused on the dust-enshrouded massive stars in M33 
identifying several candidates as extreme asymptotic giant branch 
(EAGB) stars that exhibited very red colors from \textit{Spitzer} InfraRed
Array Camera (IRAC) Band 1 3.6 $\mu$m and Band 2 4.5 $\mu$m observations, 
and are likely similar to the progenitor
of transients like SN 2008S \citep{pri08}. \citet{kha10} extended 
this effort to the nearby galaxies NGC 6946, M33, NGC 300, and M81. Of
particular interest was the identification of the brightest mid-IR
point source in M33, dubbed Object X \citep{kha11}, a likely binary 
system composed of a massive O star and a dust enshrouded red supergiant
\citep{mik15}. A similar search was also applied to 
look for $\eta$ Car-like objects in NGC 6822, M33, NGC 300, NGC 2403,
M81, NGC 247 and NGC 7793 \citep{kha13}. \citet{bri14} used 
\textit{Spitzer}
photometry to select bright candidates in the Local Group dIrr galaxies
IC 1613 and Sextans A, and their follow-up spectroscopy discovered 
five new RSGs and one yellow hypergiant candidate. 

We aim to continue to characterize the massive stellar content of 
nearby galaxies with high star formation rates starting with M83. 
M83 lies outside of the Local Group at a distance of $\sim$4.8 Mpc
\citep[$\mu \sim28.4$ mag,][]{her08,rad11}. It is a face-on galaxy 
classified as SAB(s)c \citep{dev91} with the fifth-highest H$\alpha$ 
luminosity in the local 11 Mpc$^3$ volume 
\citep[log $L_{\rm H\alpha}$ = 41.25,][]{ken08}. The star formation 
rate for a galaxy is directly proportional to its H$\alpha$ luminosity 
\citep{mur11}. M83 has also been host to six historical SNe, 
five of which have been identified as massive star or core-collapse: 
the Type IIP SN 1923A 
\citep{lam23,ros88}, the unclassified SN 1945B \citep{lil90},
the Type II SN 1950B \citep{har50,wei86}, another Type II SN1957D 
\citep{kow71,wei86}, the Type IIP SN 1968L \citep{ben68,bar79},
and the Type Ib SN 1983N \citep{tho83,por87}. The high H$\alpha$
luminosity coupled with the copious amount of historical SNe and
SN remnants \citep{bla12,bla14} imply that M83 has
a large population of massive young stars. These stars evolve
relatively quickly, meaning there may also be a large number
of dusty evolved massive stars that are prominent in the
high-quality \textit{Spitzer} archival images. We also supplement 
\textit{Spitzer}
data with ground-based near-IR and \textit{Hubble Space Telescope} 
(\textit{HST}) Wide Field Camera 3 (WFC3) observations in order to 
investigate the massive stellar content of M83.

\citet{kim12} studied the stellar content of select regions of M83
with a subset of the WFC3 data used here and came to the conclusion
that young stars are more likely to be found in concentrated aggregates
along spiral arms. \citet{lar99} reached a similar conclusion, noting
that highly crowded clusters are less than 10 Myr old, roughly the
expected age for RSGs of masses similar to that of the progenitor
of SN 2012aw. Also, \citet{cha10}, using \textit{HST} observations,
noted that M83 contained a large number of clusters with 
$M>10^{3} M_{\odot}$ and $\tau<10^{7}~yr$. We therefore focus on
methods attempting to find massive star candidates that are 
demonstrably separate in the data from cluster candidates. We also
target the spiral arm regions of M83, in order to investigate
the youngest stellar populations.

Section 2 describes the observations and data reduction, with 
some initial
analysis. Section 3 describes the combined catalog between data sets
from section 2, and our methods for
selecting massive stars from the data. In section 4, we compare
our method of selecting evolved massive stars to other methods
of photometric selection and present conclusions in section 5.


\section{Observations and Data Reduction}

\subsection{\textit{Spitzer} Observations}\label{spobs}

We extracted the mosaic image of M83 from the Local Volume Legacy 
Survey \citep[LVL,][]{dal09} in all four of the IRAC bands.
The final image analyzed
here covers an area of 15$\arcmin$ $\times$ 15$\arcmin$ with a 
pixel scale of $0\farcs75$ pix$^{-1}$ centered on the nucleus of
M83 (J2000.0: $\alpha = $13:37:00.9, $\delta = -$29:51:56).

Initial point source lists in \textit{Spitzer} IRAC 3.6 $\mu$m (Band 1)
and 4.5 $\mu$m (Band 2) bands were extracted using 
the IRAF\footnote{IRAF is distributed
by the National Optical Astronomy Observatory, which is operated by 
the Association of Universities for Research in Astronomy (AURA) Inc.,
under cooperative agreement with the National Science Foundation.} 
implementations of the DAOPHOT \citep{ste87,ste92} suite of programs.
A point-spread function (PSF) was built for each bandpass from bright,
isolated stars. Aperture photometry of the PSF stars was converted to
the Vega system using the aperture corrections and zero-point fluxes
given in the IRAC Instrument 
Handbook\footnote{
http://irsa.ipac.caltech.edu/data/SPITZER/docs/irac/iracinstrumenthandbook/
}.
PSF magnitudes for the entire set of point sources were then
calibrated by applying the offset between PSF magnitudes and
corrected aperture magnitudes for the PSF stars. The coordinates for
point sources in the 3.6 $\mu$m list were checked against those
in the 4.5 $\mu$m list using J. D. Smith's IDL code 
{\tt match\_2d.pro}
\footnote{
http://tir.astro.utoledo.edu/jdsmith/code/idl.php
}
with a half pixel search radius ($0\farcs375$). Positional coordinates
were derived from the World Coordinate System (WCS) information
contained in the header of the archival LVL image for the 3.6 $\mu$m
frame.

To eliminate the possibility of including foreground or other
previously identified objects, we checked the positions of candidates 
in our list against a number of catalogs. Mimicking the method used to
match sources between the 3.6 $\mu$m and 4.5 $\mu$m bands, we compared
our candidate positions to those in the PPMXL catalog \citep{roe10},
discarding 73 candidates having both a position match, and a proper
motion in either right ascension or declination of $>15$ mas yr$^{-1}$
(uncertainties in proper motion measurements were on the order of 10 mas). 
\citet{lar99} studied young massive star clusters in M83, publishing a
catalog of 149 clusters. Using the same half pixel search criteria, we
found matches to 5 of the \citet{lar99} clusters, and removed them from
further analysis. We also checked the cluster list from
a study of the central region of M83 by \citet{har01}, but found no
matches. \citet{her09} list 241
photometrically identified planetary nebulae in M83 from their investigation
into the mass-to-light ratio of spiral galaxy disks. A coordinate check
against this list resulted in no matches with our candidates.
In a reanalysis of European Southern Observatory (ESO) 
Very Large Telescope (VLT) imaging data covering one 
$6\farcm8$ $\times$ $6\farcm8$
field in M83, \citet{bon03} discovered 112 Cepheid variables, none
of which matched the coordinates of any of our candidates. The supernova
remnant population of M83 has been studied in some detail. We compared
\textit{Spitzer} candidates against the object lists in \citet{bla12}, 
rejecting 3 objects, and the list of new objects in \citet{bla14},
finding no matches.

The final 3778 objects in M83 are included in Table 
\ref{table:main}. The positions of sources in RA and Dec are given in 
degrees in the first two columns of the table. Based on comparison
with positions of 2MASS stars in the observed field, we estimate
the astrometry to be accurate to about $0\farcs3$.
The remaining columns are made of pairs: the source's photometry
in a bandpass followed by the uncertainty in the measurement.
The columns start at the lowest wavelength, with 
$J$, and continue with $K_{\rm s}$ (described in 
\S \ref{FourStar} for the near-IR), 3.6 $\mu$m, 4.5 $\mu$m,
5.8 $\mu$m, and 8.0 $\mu$m (described in \S \ref{combcatalog}). 
Sources with no detection in a particular 
bandpass are left with a blank entry. Some sources were only detected
in the near-IR or mid-IR and thus only have measurements in $J$ and
$K_{\rm s}$ or 3.6 $\mu$m and 4.5 $\mu$m.

We estimated the photometric completeness of the list in a
method similar to \citet{bib12}. We placed detected sources in
0.25 mag bins and fit a power law to the bright end of this distribution.
A completeness level limit of 100\% is assumed to be where this distribution
begins to drop from this power law. The 100\% completeness magnitudes 
are therefore estimated to be $[3.6]\approx18.3$ mag and 
$[4.5]\approx17.5$ mag. These limits correspond to absolute magnitudes
of M$_{[3.6]}\approx$ --10.1 mag and M$_{[4.5]}\approx$ --10.9 mag
using our assumed distance modulus of $\mu\approx28.4$ mag.
Therefore, we are only able to completely
sample the most luminous massive stars in these bands. 

\subsection{FourStar Observations}\label{FourStar}

Observations of M83 were made in $J$ and $K_{\rm s}$ with the FourStar 
instrument \citep{per13} attached to the 6.5m Baade Magellan Telescope 
at Las Campanas Observatory, on UT date 2014 Jan 2. Data reduction was
performed with the FSRED 
package\footnote{
http://instrumentation.obs.carnegiescience.edu/FourStar/SOFTWARE/reduction.html
} 
courtesy of A. Monson. The resulting $J$-band mosaic (MJD=56659.31938)
was constructed from 7 dithered exposures yielding a maximum 
exposure time of 335 seconds, while the corresponding $K_{\rm s}$ 
mosaic (MJD=56659.33165) was the
combination of 10 frames giving a maximum exposure time of 582 seconds.
Astrometric calibration employed SWarp \citep{ber02} software
and stars from the 2MASS \citep{skr06} catalog. The final images have
a field of view of 11$\arcmin$ $\times$ 11$\arcmin$ centered on the
nucleus of M83 with a pixel scale of $0\farcs16$ pix$^{-1}$. 

Photometry for the $J$ and $K_{\rm s}$ bands was conducted in the same
fashion as for the \textit{Spitzer} data. The PSF varied across the four
CCDs making up the FourStar instrument, so several PSF stars in each 
quadrant of the final mosaic were selected. The PSF stars were chosen from
the 2MASS catalog and used for both aperture correction and 
photometric calibration. Because the photometric standard stars
were on the same frame as the candidate objects, the transformation
equations simplified to a zero-point for each band and color term 
between the two bands. The computation of transformed magnitudes thus
required a source to have a measurement in both the $J$ and $K_{\rm s}$
bands within a matching radius of 2 pixels corresponding to 
$\sim0\farcs32$. The 2MASS stars in each bandpass were also
subjected to the same transformation, and the comparison of our 
measured values and those published in the 2MASS catalog are shown in
Table \ref{2masscompare}. The average difference in the $J$-band
is $0.047\pm0.039$ mag, while the average difference for the $(J-K_{\rm s})$ 
color is $0.056\pm0.045$ mag. The average difference in $J$ is
nearly insignificant, however, the similar difference in the
$(J-K_{\rm s})$ color may indicate some systematics with the $J$-band
measurements.

Artificial star tests were used to estimate the photometric
uncertainty of a measurement similar to the method employed by
\citet{gra13}. Using the PSF constructed for each bandpass
image, we added one thousand artificial stars (a number roughly
equal to the total number of point source detections matched between
bands) to each frame via the ADDSTAR routine in the DAOPHOT package.
We then performed photometry on these targets in the same manner
used on the point source candidate list and repeated this
process three times. We computed the root-mean-square (rms) 
of the differences between input and output magnitudes for 0.2 mag bins. 
This rms difference was then added in
quadrature with measurement uncertainties from ALLSTAR falling into 
that particular bin, to make the final estimate of measurement
uncertainty. It should be noted that this rms difference was typically
very close in value to the uncertainty estimates from ALLSTAR.
The final point source list of 975 objects are included in Table 
\ref{table:main}. Sources detected in the near-IR 
are sorted by RA in column 1, with column 2
listing the declination, both in degrees. Again, from comparison
with 2MASS sources, we estimate that the FourStar astrometry is
accurate to roughly $0\farcs3$.
Columns 4 through 7 correspond to $J$ 
magnitude and measurement uncertainty, and $K_{\rm s}$ magnitude 
and measurement uncertainty for sources detected in the near-IR.
For sources with no $J$ and $K_{\rm s}$ measurements these columns 
are left blank. We did not correct the point source photometry
for foreground Galactic extinction. However, for completeness, the
current best estimate using the \citet{sch11} infrared-based dust map
gives 0.047 and 0.020 mag in $J$ and $K_{\rm s}$, respectively. 

Using the same method as described in \S\ref{spobs}, we found the
100\% completeness levels to be 18.0 mag (M$_J\approx$ --10.4) in $J$ 
and 16.6 mag (M$_{K_{\rm s}} \approx$ --11.8) in $K_{\rm s}$.

\subsection{$HST$ WFC3 Observations}

\citet{bla14} studied supernova remnants in M83 with seven fields
of \textit{Hubble Space Telescope} (\textit{HST}) Wide Field Camera 3 (WFC3) 
observations in multiple bandpasses. Two fields come from the
Early Release Science Program (ID 11360; R. O'Connell, PI) with
the remaining five coming from the cycle 19 \textit{HST} General Observer
program 12513 (W. Blair, PI).
These seven fields cover the majority of the bright disk region
of M83. For the reduction and discussions concerning the 
photometry of these data, we direct the reader to 
\citet[][and references therein]{bla14}. We specifically used
the imaging and photometry in the F336W (Johnson $U$), F438W (Johnson $B$),
F555W (Johnson $V$) or F547M (Str\"omgren $y$, easily 
converted to Johnson $V$), and F814W (Johnson $I$) bands.
The quality of the astrometry for the $HST$ data are better
than $0\farcs1$ \citep{bla14}.


\section{Results}

\subsection{Combined Catalog of Point Sources}\label{combcatalog}

In order to characterize the point sources, sources detected
in both the near-IR and mid-IR were matched by position as described
above. Final
coordinates were adopted from the shortest wavelength, $J$,
owing to the band having the better spatial resolution compared to
longer wavelength bands. To supplement these measurements, 
aperture photometry using the IRAF task APPHOT/PHOT was
performed on the remaining IRAC bands, 5.8 $\mu$m and 8.0 $\mu$m
at the pixel positions of sources having $J$, $K_{\rm s}$,
3.6 $\mu$m and 4.5 $\mu$m photometry. The aperture photometry in
these two bands was calibrated in the same way as described
for the PSF stars in the 3.6 $\mu$m and 4.5 $\mu$m bands in
the previous section, with appropriate constants applied
from the IRAC instrument handbook.

This near- and mid-IR combined catalog
of 221 point sources in M83 is presented in Table \ref{table:main}.

\subsection{Candidate Massive Star Collection}

Identifying massive star candidates among our point source
catalog presents several challenges. For example, 
the average FWHM of objects in the 3.6 $\mu$m band
is $\sim1\farcs7$, which corresponds to 40 pc at the 4.8 Mpc
distance to M83. Thus, virtually all the clusters identified
in \citet{lar99} will be unresolved on ground-based images, 
leaving open the possibility of confusing massive stars 
with compact clusters or unresolved associations. 
We also expect contamination from 
foreground Milky Way halo objects like M-dwarfs, and background
galaxies. One way to 
overcome this limitation is to compare the location of objects 
on various color-magnitude diagrams (CMDs) and other photometric
diagnostic plots. Another way to address this issue is to use
higher resolution \textit{Hubble} images, as employed in the current 
paper.

\subsubsection{Mid-IR Selection Criteria}

We applied mid-IR criteria to select massive stars from our
sample following the \textit{Spitzer Space Telescope} Legacy Survey 
known as ``Surveying
the Agents of a Galaxy's Evolution'' (SAGE) studied both the 
LMC \citep{mei06} and SMC \citep[SAGE-SMC;][]{gor11} 
allowing a characterization of
the IR properties of the massive stellar content of both
galaxies \citep{bon09,bon10}. In both the LMC \citep{bon09}
and SMC \citep{bon10}, massive, evolved stars with previously known 
spectral types were shown to group in certain regions of the [3.6] versus 
[3.6]--[4.5] CMD, thanks to the well-studied
nature of the brightest stars of these nearby galaxies. 
Specifically, looking at Figure 2 of both \citet{bon09} and
\citet{bon10}, RSGs are clumped in a region
where $-0.3<[3.6]-[4.5]<0.0$ and $-13.0<M_{[3.6]}<-9$, 
while supergiant Be (sgB[e]) stars are located in a band further to the red, 
with [3.6]$-$[4.5] between 0.6 and 0.9 and $-13<M_{[3.6]}<-9$. 
RSGs appear ``blue'' owing to the suppression of flux in [4.5] from the 
CO and CO$_2$ molecular bandheads \citep{ver09}.
LBVs occupy the [3.6]$-$[4.5] color space between these two
groups. There are always exceptions to these generalizations, 
and one concern may be that the cuts for RSGs may not include 
stars with dusty envelopes. There is evidence that these
photometric cuts are leaving out RSG. For example, in \citet{bri14},
the RSG IC 1613 1 was originally thought to be an LBV based on
its color: [3.6]$-$[4.5]$=0.67$. Follow up spectroscopy, however,
revealed it to be an early M supergiant.

Figure \ref{3p6vcolor} shows the [3.6] versus [3.6]$-$[4.5]
CMD of point sources from Table \ref{table:main}. Sources are
plotted as black dots except for the two regions empirically shown
to contain RSGs and sgB[e] stars as discussed above. The RSG
region is outlined in a red box, and point sources are shown with 
red dots, while the sgB[e] region follows the same
prescription, but with the color blue.
Applying these regional criteria for massive star candidates yields 638 
RSGs and 363 sgB[e] candidate stars. Also plotted in Figure \ref{3p6vcolor}
are prominent massive stars 
from the literature: $\eta$ Car \citep{hum94}, M33 Variable A
\citep{hum06}, and Object X 
\citep{kha11}. To
represent the accuracy of the photometry, we have plotted ellipses
on the left side showing the typical uncertainties in [3.6] 
and [3.6]$-$[4.5] for each one magnitude bin.

One major difference between the LMC and SMC versus M83 is
that background extragalactic sources become more numerous in the mid-IR CMD
at the distance of M83. \citet{ash09} showed that for the \textit{Spitzer}
Deep, Wide-Field Survey (SDWFS), galaxies make up most of their
detections. This is especially true past $[3.6]>16$ mag, where
many of our massive star candidates exist. We determined the
approximate contamination by extracting 42,465 sources from the final
version of the SDWFS catalog \citep{koz10} from a randomly
selected one square degree area. These are plotted in Figure 
\ref{3p6vcolor-all} as small grey dots. Figure \ref{3p6vcolor-all} is
the same as Figure \ref{3p6vcolor} except it 
also shows the possible regions of contamination 
by background sources from the SDWFS catalog. This is not
entirely representative of the expected contamination in the
field of M83, as it contains all objects within a one square 
degree region while the analyzed M83 \textit{Spitzer} image covers only  
15$\arcmin$ $\times$ 15$\arcmin$. 

To further explore the background contamination, the same
selection criteria used to find massive star candidates in M83 
were applied to the sample from the SDWFS, yielding 4309
objects in the RSG area and 5348 objects in the sgB[e]
region. We then scaled these numbers for the area encompassed by
the extracted \textit{Spitzer} image of M83, 15$\arcmin$ $\times$ 
15$\arcmin$ or 0.0625 square degrees. The final estimates for
contamination are 270 (42\%) in the RSG region, leaving 368 RSG
candidates and 335 (92\%) in the sgB[e] region, with only 28 objects
not likely to be background contaminants. \citet{kha13} used similar
methods to estimate their contamination, finding it to be much
less for galaxies closer to the Milky Way. In a study of the asymptotic
giant branch (AGB) and super-AGB (SAGB) population of the LMC,
\citet{del14} showed that the most massive 
($6-8M_{\odot}$) AGBs, the OH/IR stars, exist at
luminosities above the classic limit for AGBs of 
$L \sim 5 \times 10^4 L_{\odot}$ or $M_{bol} = -7.1$ mag. The AGB region
outlined in \citet{kha10} lies in a slightly reddened part of the
CMD, redward of [3.6]$-$[4.5]$=$0.1 mag and where $M_{\rm [3.6]} < -10$ mag.
However, given the photometric uncertainties at the distance of
M83, there is the possibility that some photometrically identified
evolved massive star candidates may actually be the most 
massive and most luminous AGB/SAGB stars.

The strongest massive star candidates from the \textit{Spitzer} photometry
are those objects that lie outside the background contaminated
region shown in Figure \ref{3p6vcolor-all}. 
These include a number of objects that may be very
red objects [3.6]$-$[4.5] $> 1.5$ mag or objects with colors bluer
than RSGs. The strongest RSG candidates lie in the region bounded by 
$-0.3 <$ [3.6]$-$[4.5] $< -0.1$ mag for 15.4 $<$ [3.6] $<$ 16.5 mag
and in the bounding box above the line connecting the two points
of [3.6] = 16.5 mag and [3.6]$-$[4.5]$=-0.1$ mag with [3.6] = 18.0 mag
and [3.6]$-$[4.5]$=-0.3$ mag. It should be noted that photometric 
uncertainties may shift some candidates into and other objects out of
this region. Regardless, these criteria resulted in the 
identification of 118 candidates from the \textit{Spitzer} photometry. 
Further refinement was made via plotting the positions
of these candidates over an image of the galaxy. The extent of the images
used in this analysis (15$\arcmin$ $\times$ 15$\arcmin$) goes well
beyond the disk light of M83. While massive stars exist in the
outer regions of M83, they will be more numerous particularly
compared to background galaxies in the regions of disk light
of M83. Because we want to maximize our potential for selecting
massive star candidates for spectroscopic follow-up, we therefore
removed detected sources lying outside of a square region centered 
on the nucleus of M83 of 800 pixels or 10\arcmin. This region
is selected based on the extent of the disk of M83 as seen in
the 3.6 $\mu$m image. The remaining RSG candidates within this
region are 68 point sources
selected only from \textit{Spitzer} 3.6 $\mu$m and 4.5 $\mu$m observations 
and denoted by ``\textit{Spitzer} RSG Candidate'' in column 15 of 
Table \ref{table:main}.

\subsubsection{Near-IR Selection Criteria}

In order to expand the massive evolved candidate star list, we
used the FourStar data in $J$-band along with the \textit{Spitzer} 
3.6 $\mu$m measurements to construct a plot of [3.6] versus 
$J-$[3.6] similar to Figure 3 in \citet{bon09}. 
Figure \ref{jm3p6v3p6} shows our version of the plot. We have
regions of interest outlined in red for the location of
spectroscopically known red supergiants in the LMC \citep{bon09},
1.0 $<$ $J-$[3.6] $<$ 2.0 and fainter than $M_{\rm [3.6]}=-12$,
while the blue outline shows the location of known sgB[e] stars,
2.0 $<$ $J-$[3.6] $<$ 4.5 and fainter than $M_{\rm [3.6]}=-12$.
A total of 148 candidates were found in these regions. At the risk
of removing possible interesting massive star candidates
(massive stars outside of this region due to binary interactions
or SN kicks), but
increasing our chances of selecting individual massive stars,
we again chose to exclude any objects lying outside a
box of 10\arcmin~centered on the nucleus of M83.
When combined with those selected from our analysis of \textit{Spitzer}
observations 31 sources were selected using both sets of criteria, 
leaving 185 massive star candidates. 

\subsubsection{High Angular Resolution \textit{HST}/WFC3 
Optical Selection Criteria}

Cross referencing coordinates of the 185 candidates
selected from near- and mid-IR photometric criteria
with positions in the WFC3 images, we visually 
inspected the area surrounding each 
candidate and classified it based on its appearance. We ranked
each candidate based on the probability of confirming their
massive nature via spectroscopic follow up of the target. 
Rank 1 is used for
objects that are the most isolated, with few nearby stars of
comparable brightness. Those objects with slightly more
crowded fields, but maintaining the appearance of a single star
are put into rank 2. In rank 3, fields get more crowded, making
identification of the individual objects more difficult. Some
objects classified as rank 3 show a diffuse nature. For objects in 
rank 4, crowding was worse than
in rank 3 or objects are even more difficult to recover. Rank 5 is reserved
for the worst cases of crowding, diffuse nature of the primary
candidate, or a complete lack of recovered object at the position.

The WFC3 fields do not cover all of the disk of M83. Because of this
only 112 of the 185 massive star candidates may be checked by visual inspection.
This resulted in the identification of 55 ($49\%$) candidates with 
ranks 1 and 2 for best suited for spectroscopic follow-up observations. 

Figure \ref{HSTimages} shows examples of each
rank, 1 through 5, while the online figure set shows a 
postage stamp for each of the 112 candidates grouped by rank. 
These ranks, as well as a short comment
to explain the rank for each candidate, are given in Table 
\ref{maintable} for all 185 candidates selected from IR data. 
Objects not in the \textit{HST} field of view have a comment stating
as much. Also listed
are broadband photometric measurements and uncertainties 
in Johnson $U,B,V$ and $I$ as well as the available near- and mid-IR 
photometric measurements. The uncertainties of the \textit{HST} optical
measurements are merely the output from DAOPHOT and have not been 
more closely checked. They are listed 
here to give the reader a general idea of the accuracy of the \textit{HST}
photometry. The magnitudes listed in Table \ref{maintable} 
are not corrected for either Galactic extinction, nor reddening 
internal to M83. However, for reference, Galactic extinction is 
$A_U = 0.288$, $A_B = 0.241$, $A_V = 0.182$, $A_I = 0.1$ mag
from the \citet{sch11} IR-based dust map, and the average M83 
internal extinctions from \citet{kim12} based on \textit{HST} observations
of the central region are
$A_U = 0.696$, $A_B = 0.559$, $A_V = 0.441$, $A_I = 0.256$ mag.
To illustrate the locations of these candidate RSGs, Figure \ref{m83pic} shows
the greyscale IRAC Band 1 3.6 $\mu$m image used in the analysis with the 
sky positions of the candidates overplotted. Candidates of the
best rank (ranks 1 and 2) are plotted as red circles, 
while those of lower quality (ranks 3 through 5) are 
represented by blue circles.


\section{Discussion}

There have been several surveys
for massive stars that have used broadband optical photometry 
\citep[e.g.][]{mas98,lev12,gra13} as a principal means to find
red, blue, and yellow supergiants. Thus, a 
comparison of optical versus IR selection techniques is worthy of
a brief discussion.

The primary reason to utilize mid-IR observations to explore the
massive stellar content of a galaxy is that evolved massive stars,
with substantial dust or circumstellar material, will be more easily
detectable in the IR as opposed to the optical. Deep optical photometry
of galaxies has not been conducted in a systematic fashion 
beyond the Local Group Galaxy Survey effort of \citet{mas06}. Plus,
mid-IR selection criteria have been shown successful in the Local Group
after spectroscopic follow up \citep{bri14}.

One drawback with the current mid-IR surveys is the difference
in resolution compared to near-IR and optical images. 
Massive stars are born, and mostly die, in clusters
or associations, meaning spatial resolution at the distance of M83
is a prominent concern. The PSF of the near-IR observations is just
below 1$''$ corresponding to a size of 23 pc at M83's
distance. This size is
smaller than the 40 pc size of the \textit{Spitzer} PSF, but may still 
easily include clusters, associations or several stars. 
OB associations can
have half-light radii of a few parsecs and extend over several
tens of parsecs, while young massive clusters can exist in a 
volume less than one cubic parsec \citep{cla05}. One such example
is the cluster Stephenson 2, which contains roughly 30 RSGs in
a radius between 3.2 and 4.2 pc \citep{neg12}.
The regions of M83 covered by the seven \textit{HST} fields
have an order of magnitude better resolution than the FourStar
observations. This PSF corresponds to $\sim$0$\farcs$1 or just over
2 pc. While some objects may still be compact clusters, we can
eliminate many objects that are larger than 2 pc in size,
increasing our odds of observing only individual stars with
follow-up spectroscopy. 

The second issue with using mid-IR data concerns the limiting
magnitude attainable. We found absolute magnitude limits of
M$_{[3.6]}\simeq-10.1$ mag for the completeness of the survey, meaning
we are completely sensitive to only the brightest RSGs at the
distance of M83. Similarly, \citet{kha10} noted that for NGC 6946
at a distance of 5.6 Mpc, the 3$\sigma$ detection limit in the
4.5 $\mu$m band corresponded to an absolute magnitude of close to $-10$. 
Therefore, given the limits in both photometry and resolution,
mid-IR studies intended to study the stellar content of galaxies 
beyond $\sim$5 Mpc with current archival data from \textit{Spitzer} will
be severely limited.

Ideal candidates are those that are isolated
and meet photometric criteria in the IR that have been previously 
successfully used to identify massive stars. 
Because we have deep optical photometry (from the \textit{HST} images) 
for our IR selected candidates, 
we can explore the differences and similarities in selecting 
candidates using IR photometry versus selecting candidates using
optical photometry. 

In pioneering the effort to find red supergiants in nearby galaxies,
\citet{mas98} showed that while both the $V-R$ and $B-V$ photometric
colors change with
effective temperature, $B-V$ also changes with surface gravity, 
and can therefore be used to uniquely identify RSGs from foreground,
Milky Way red halo stars. A similar approach was implemented by \citet{gra13}
with $B-V$ and $V-I$ colors. \citet{gra13} extracted a theoretical 
supergiant sequence
from the stellar models of \citet{ber94} to further illustrate the 
usefulness of such a color-color diagram in selecting massive star
candidates. Ten of the 112 candidates visually inspected
in the \textit{HST} images were not recovered in one or more of the filters
needed for plotting on this color-color diagram. 
Figure \ref{bvvsvi} shows the 102 (of 112) 
candidates listed in Table \ref{maintable} with $B$, $V$, 
and $I$ magnitudes on a color-color diagram of $B-V$ versus $V-I$. 
In total, there are
49 candidates with rank 2 or lower plotted in red in Figure \ref{bvvsvi}
and 53 candidates with a rank higher than 2 plotted in black.
Of the 49 highly ranked visually inspected candidates, 
24 (49\%) lie in the region of the color-color diagram
similar to the RSG candidates of \citet{gra13}. Within uncertainties,
another four strong candidates also lie within this region. 
Based on the location of our candidates on
the similar plot of $B-V$ versus $V-I$, we expect that $\sim$50\%
of the strong (49 red points in Figure \ref{bvvsvi}) candidates 
we selected via near- and mid-IR photometry are
truly RSGs. To be certain, we need follow-up spectroscopy to confirm
this hypothesis. For comparison, \citet{mas98}
applied a photometric cutoff in the $B-V$ versus $V-R$ colors, 
then performed spectroscopic follow-up 
observations in order to confirm the objects as RSGs and found that for
the faintest candidates, the success rate of the optical photometric criteria 
in selecting RSGs was 82\%. 
To compare our expected success rate of recovering
RSGs with that from optical surveys, we must note a few caveats.
First, the optical photometry used by \citet{mas09}, for example, is
much more accurate than the photometry from the \textit{HST} data. This is
simply due to the much greater distance to the stars in M83 than to 
the studied stars in Local Group galaxies. The second caveat relates
also to the accuracy of the photometry, this time from \textit{Spitzer}. 
Selection criteria are only as good as the accuracy of the photometry, and
as can be seen in Figures \ref{3p6vcolor} and \ref{3p6vcolor-all},
within uncertainties, some RSG candidates may move in and out of
the region encompassed by our selection criteria.
Because of these caveats, our predicted RSG success rate is lower 
than that from optical observations. 

To aid in illustrating where confirmed RSGs reside, 
in Figure \ref{bvvsvi} we have plotted 
spectroscopically confirmed RSGs (grey dots) from the Local Group
that have $B$, $V$, and $I$-band measurements: 119 from M31 that
are not labeled as ``crowded'' objects \citep{mas09}, 46 from M33
that are not ``multiple'' \citep{mas06}, 73 from the Milky Way
\citep[][and references therein]{duc02,lev05}, 41 from the LMC
\citep{bon09}, 19 from the SMC \citep{bon10}, and 30 from the Dwarf
Irregular galaxies (dIrrs) in the Local Group 
(\citealt{lev12,bri14}; Britavskiy et al. in prep.). The
box outlines the region described in \citet{gra13}
where \citet{bar06} show that cluster candidates exist. Interestingly,
this region also is predominantly the locus of the 
massive stars such as sgB[e] from \citet{bon09,bon10} (blue dots), 
and the list of LBVs, LBV candidates, \ion{Fe}{ii} emission line
stars, and other (blue) supergiants in M31 and M33 from 
\citet{hum14} shown as green dots. Comparing our candidates with
the locations of the spectroscopically classified massive stars
from the literature, objects in the upper right of the plot are
most likely RSGs, while those in the lower left are their bluer massive
star counterparts. The average reddening \citep{kim12} is shown as an arrow,
meaning some candidates outside the described regions are simply
reddened RSGs in the upper part of the plot, or reddened sgB[e] stars 
(for example) in the lower central part of the plot. The
strong candidates (red dots) in the lower right part of the plot are not
very red in $B-V$ while being quite red in $V-I$. \citet{gra13}
put forth the explanation that objects in the bottom right of the 
color-color diagram may contain blended objects or compact 
\ion{H}{ii} regions. Indeed, at the limit of the \textit{HST} resolution,
ultra compact \ion{H}{ii} regions, which are typically connected
to small clusters of stars \citep{con08}, may be contaminants.
In a study of the clusters in the central fields of the \textit{HST}
data for M83, \citet{cha10} show that 
reddening may cause the $V-I$ values of candidate LBVs
to move out of where they are expected in the same
way as seen in our plot. \citet{whi11} 
and \citet{kim12} bring up the possibility that these objects
may be the chance superposition of a red and a blue star. 
Spectroscopic follow-up is the only way to definitively 
determine the nature of these objects as either reddened LBVs, 
a superposition of two different kinds of stars or simply clusters
of stars. It should also be mentioned that \citet{mas09}
discussed that 5 of 19 spectroscopically confirmed RSGs
in M31 from \citet{mas98} would not have met their photometric criteria. 
They stressed
that RSGs may exist in other regions of the color-color diagram,
but reiterated the point that their optically photometric criteria have
a superb success rate of $>82$\% \citep{mas98} at selecting RSGs
based on confirmation from follow-up spectroscopy. Again, this is higher than
our expected $\sim$50\% success rate based on IR photometry. 

How does selecting evolved massive stars via IR photometry compare
with the \citet{mas98} optical selection method? Plotted in 
Figure \ref{bvvsvilocal} by color and galaxy are the 
spectroscopically confirmed RSGs shown in grey in Figure 
\ref{bvvsvi}. Also plotted is the supergiant sequence from
\citet{gra13} and \citet{ber94} for reference. Because optical
surveys were performed first and are thus more numerous and typically 
have better resolution,
a very small number of confirmed RSGs come from the IR-selected 
candidates in dIrrs (\citealt{bri14}; Britavskiy et al. in prep.),
which are shown as red triangles outlined in black.
The spectroscopically confirmed IR-selected candidates in the
dIrrs in the Local Group lie in exactly the same region 
as those spectroscopically confirmed RSGs selected via 
optical photometry, and are represented by downward pointing
red triangles that are not outlined in black. Clearly,
there are too few RSGs selected via the IR to draw strong conclusions,
but these first few attest to the usefulness of selecting RSGs 
from IR observations. Looking further at the plot,
there are a few interesting possibilities that will be confirmed
or rejected with more observations. 
For the confirmed RSGs in the SMC in 
Figure \ref{bvvsvilocal}, are the six RSGs with lower $B-V$ 
representative of a metallicity effect? These RSGs are all
K-type, and the trend of earlier average spectral type of RSGs
with decreasing metallicity of the host galaxy has been shown
by \citet{lev12}. More observations of RSGs in metal-poor 
galaxies are needed to confirm or refute this possible trend. 
The Local Group dIrr 
RSGs do not yet have deep enough photometry to
indicate if this trend is real. Another question raised by this
plot is why do the LMC RSGs, Milky Way RSGs, and M83 RSGs suffer
from so much scatter on this plot?

In Figure \ref{bvvsvisptype} we plot RSGs grouped by four spectral 
types: early K, late K, early M, and late M. The K spectral type
RSGs appear less reddened, and show less scatter. This follows from 
the intrinsic bluer colors of K-type RSGs compared to M-types.
However, K-type RSGs also appear to have less reddening than their
later-type companions. This is likely due to the fact they come
from, on average, lower metallicity galaxies. 
One interesting possibility for this scatter for the later-type 
RSGs may be related to
the behavior of HV 11423 in the SMC. \citet{mas07} tracked HV 11423
through changes in its spectral type from K0 I to M5 I. The star
changed by 2 mag in $V$, with changes in spectral type attributed
to both the change in effective temperature, and the creation 
and dissipation of dust \citep{mas07}. HV 11423 is an exceptional
case, but it would be interesting to investigate via long
term observations whether 
the scatter of later spectral types in Figure 
\ref{bvvsvisptype} is connected with a cycle of production and 
destruction of dust.

What kind of RSGs do we expect to find in M83? The average spectral
type for RSGs in the Milky Way is M2 \citep{lev12}. The disk
metallicity of M83 was shown to be 1.9$\times$ Solar by \citet{gaz14}.
Thus, we may expect the average spectral type of an M83 disk RSG
to be M2 or later. In addition, we may expect the typical M83 RSG
to experience greater mass loss rates, as mass loss rates
tend to increase with later spectral type.


\section{Conclusions}

We present point source catalogs for a 15$\arcmin$ $\times$ 
15$\arcmin$ region centered on M83 based on \textit{Spitzer}
mid-IR and near-IR photometry. From these catalogs,
we have selected candidate massive stars based on
their mid- and near-IR photometric properties. These candidates have
been culled by checking against catalogs of known objects, and
also by inspection on high resolution \textit{HST} images. The remaining 49
objects are strong candidates for evolved massive stars, and await 
follow-up spectroscopy to further determine their nature and
quantify the success rate of this technique in detecting 
evolved massive stars.

We make the full catalog of point sources available to
the community in order, for example, for future researchers 
to characterize the progenitors of core-collapse SNe.
The methodology used in this paper may be readily applied to
other nearby ($\leq5$ Mpc) galaxies to investigate their evolved massive
stellar populations. High quality archival \textit{Spitzer} images 
exist for all nearby ($\leq10$ Mpc) galaxies courtesy of the 
SINGS and LVL surveys \citep{ken03,dal09}. Archival 
\textit{HST} images also exist for many of the nearby galaxies. 
Analysis of follow-up spectroscopy of candidates discussed here
for M83 will be needed to further test this methodology. However,
studies starting merely with \textit{Spitzer} photometry have
already proved to be successful in discovering evolved 
massive stars (\citealt{bri14}; 
Britavskiy et al. in prep.). The validation of the
usefulness of this IR method of determining massive star candidates
is important, especially because in many cases, deep optical
photometry simply does not exist. This method may also
provide an alternative
to the previously employed optical methods to investigate
the massive stellar content of the Local Universe.


\begin{acknowledgements}

We thank the referee, I. Negueruela for comments, suggestions,
and a keen eye that helped improve the content, flow and 
presentation of the paper. We also thank the editor, Rubina Kotak
for further comments and suggestions that have helped to clarify
the discussion in the introduction.
SJW and AZB acknowledge funding by the European Union
(European Social Fund) and National Resources under the
``ARISTEIA'' action of the Operational Programme ``Education
and Lifelong Learning'' in Greece. Support for JLP is provided
in part by the Ministry of Economy, Development, and Tourism's
Millennium Science Initiative through grant IC120009, and 
awarded to the Millennium Institute of Astrophysics, MAS.
We wish to thank Rubab Khan
and Kris Stanek for discussions and guidance concerning \textit{Spitzer}
photometry. We also thank Andy Monson for helping with FourStar
data reduction. We extend our gratitude to the LVL Survey for making
their data publicly available.
This publication makes use of data products from the Two Micron All
Sky Survey, which is a joint project of the University of 
Massachusetts and the Infrared Processing and Analysis 
Center/California Institute of Technology, funded by the National 
Aeronautics and Space Administration and the National Science Foundation.
This research has made use of the VizieR catalogue access tool, 
CDS, Strasbourg, France. This work is based [in part] on observations 
made with the
\textit{Spitzer Space Telescope}, which is operated by the Jet
Propulsion Laboratory, California Institute of Technology, under
a contract with NASA.

\end{acknowledgements}



\bibliographystyle{aa}
\bibliography{aa.bib}

\clearpage

\begin{table*}

\centering

\caption{Infrared Photometry of 4528 Point Sources in the field of M83}
\label{table:main}

\begin{tabular}{ccccccccccccc}
\hline\hline
RA(J2000) & Dec(J2000) & $J$     & $\sigma_J$ & $K_{\rm s}$ & $\sigma_{K_{\rm s}}$ & [3.6] & $\sigma_{3.6}$ & [4.5] & $\sigma_{4.5}$ & [5.8] & $\sigma_{5.8}$ & ... \\
  (deg)   &    (deg)   &  (mag)  &  (mag)     &  (mag)      &  (mag)               & (mag) &         (mag)  & (mag) &  (mag)         & (mag) &  (mag)         & ... \\
\hline					     	           	                  	         			            		              	
204.16129 & --29.85840 &  16.40  &   0.02     &  16.01      &   0.03               & 15.91 &          0.03  & 16.03 &   0.07         & 14.90 &   0.07         & ... \\
204.16831 & --29.90027 &  17.42  &   0.03     &  16.52      &   0.03               & 17.02 &          0.04  & 17.27 &   0.11         & 17.09 &   0.28         & ... \\
204.17050 & --29.81191 &  16.64  &   0.02     &  16.27      &   0.05               & 16.47 &          0.05  & 16.60 &   0.06         & 16.21 &   0.11         & ... \\
204.17440 & --29.86086 &  18.31  &   0.05     &  16.65      &   0.06               & 16.73 &          0.06  & 16.99 &   0.08         & 13.31 &   0.01         & ... \\
204.18935 & --29.91848 &  16.28  &   0.01     &  15.86      &   0.02               & 15.91 &          0.04  & 16.07 &   0.05         & 15.01 &   0.04         & ... \\
204.20567 & --29.92462 &  17.81  &   0.04     &  16.80      &   0.05               & 16.77 &          0.05  & 17.01 &   0.07         & 16.31 &   0.25         & ... \\
204.21006 & --29.91604 &  18.40  &   0.05     &  16.71      &   0.05               & 16.85 &          0.05  & 17.08 &   0.07         & 13.79 &   0.03         & ... \\
204.21704 & --29.83906 &  17.41  &   0.03     &  16.40      &   0.04               & 16.70 &          0.05  & 16.84 &   0.05         & 13.46 &   0.17         & ... \\
204.22253 & --29.82444 &  17.43  &   0.05     &  15.97      &   0.06               & 15.55 &          0.04  & 15.66 &   0.06         & 15.92 &   0.76         & ... \\
204.23417 & --29.94974 &  19.66  &   0.15     &  16.94      &   0.10               & 16.37 &          0.03  & 16.50 &   0.06         & 15.82 &   0.11         & ... \\
\hline
\end{tabular}
\begin{center}
This table is available in its entirety in a machine-readable form
in the online journal. A portion is shown here for guidance 
regarding its form and content.
\end{center}
\end{table*}

\begin{table*}

\centering

\caption{FourStar Photometry Compared to 2MASS Photometry}
\label{2masscompare}

\begin{tabular}{ccccccc}
\hline\hline
2MASS ID & $J_{\rm PSF}$ & $J_{\rm 2MASS}$ & Difference & $(J-K_{\rm{s}})_{\rm FourStar}$ & $(J-K_{\rm{s}})_{\rm 2MASS}$ & Difference \\
 & (mag) & (mag) & (mag) & (mag) & (mag) & (mag)\\
\hline
13372193--2947529 & 15.761(10)  & 15.694(58)~~ & ~~0.067 & 0.889(23) & 0.888(141)  & ~~0.001\\
13371693--2947174 & 16.105(8)~~ & 16.067(88)~~ & ~~0.038 & 0.896(21) & 0.850(188)  & ~~0.046\\
13371216--2948184 & 16.235(15)  & 16.192(99)~~ & ~~0.043 & 1.027(37) & 0.923(195)  & ~~0.104\\
13364314--2947289 & 15.781(12)  & 15.872(77)~~ & --0.091 & 0.981(46) & 0.947(145)  & ~~0.034\\
13364385--2949437 & 15.427(11)  & 15.448(63)~~ & --0.020 & 0.767(42) & 0.634(130)  & ~~0.133\\
13363909--2949390 & 15.696(7)~~ & 15.729(77)~~ & --0.033 & 0.284(28) & 0.242(226)  & ~~0.042\\
13363400--2949388 & 16.011(12)  & 16.000(82)~~ & ~~0.011 & 0.708(41) & 0.852(164)  & --0.144\\
13363098--2950437 & 15.155(10)  & 15.194(48)~~ & --0.039 & 0.850(36) & 0.806(91)~~ & ~~0.044\\
13363542--2953303 & 16.219(10)  & 16.366(130)  & --0.147 & 0.993(24) & 0.950(250)  & ~~0.043\\
13370045--2957177 & 15.670(7)~~ & 15.712(65)~~ & --0.042 & 0.487(25) & 0.463(178)  & ~~0.024\\
13370527--2956231 & 15.962(10)  & 15.939(80)~~ & ~~0.023 & 1.076(26) & 1.057(146)  & ~~0.019\\
13372091--2949532 & 15.234(9)~~ & 15.249(47)~~ & --0.015 & 0.844(19) & 0.884(89)~~ & --0.040\\
\hline
\end{tabular}
\end{table*}

\begin{table*}

\centering

\caption{Massive Evolved Star Candidates in M83 from $HST$, Near-IR, and $Spitzer$ Photometry}
\label{maintable}

\begin{tabular}{ccccccccccccc}
\hline\hline
RA(J2000) & Dec(J2000) & $U$ & $\sigma_U$ & $B$ & $\sigma_B$ & $V$ & $\sigma_V$ & $I$ & $\sigma_I$ & $J$ & ... \\
(deg) & (deg) & (mag) & (mag) & (mag) & (mag) & (mag) & (mag) & (mag) & (mag) & (mag) & ... \\
\hline
204.27068 & --29.78285 &   . . . &   . . . &   . . . &   . . . &   . . . &   . . . &   . . . &   . . . & 18.07   &  ... \\
204.27253 & --29.78501 &   . . . &   . . . &   . . . &   . . . &   . . . &   . . . &   . . . &   . . . &   . . . &  ... \\
204.17782 & --29.78529 &   . . . &   . . . &   . . . &   . . . &   . . . &   . . . &   . . . &   . . . &   . . . &  ... \\
204.33232 & --29.78565 &   . . . &   . . . &   . . . &   . . . &   . . . &   . . . &   . . . &   . . . &   . . . &  ... \\
204.25152 & --29.80064 & 26.67   &    0.37 & 26.56   &    0.18 & 24.08   &    0.07 & 20.78   &    0.03 & 18.74   &  ... \\
204.24779 & --29.80129 & 26.90   &    0.42 & 25.79   &    0.13 & 23.43   &    0.05 & 20.68   &    0.03 & 18.46   &  ... \\
204.27008 & --29.80718 & 25.25   &    0.19 & 24.72   &    0.08 & 22.35   &    0.04 & 20.21   &    0.03 & 18.28   &  ... \\
204.31808 & --29.80785 &   . . . &   . . . &   . . . &   . . . &   . . . &   . . . &   . . . &   . . . &   . . . &  ... \\
204.27208 & --29.80829 &   . . . &   . . . & 29.69   &    0.90 & 26.68   &    0.22 & 22.04   &    0.04 & 18.94   &  ... \\
204.23951 & --29.80921 & 26.16   &    0.30 & 27.44   &    0.27 & 25.65   &    0.13 & 21.79   &    0.04 & 18.66   &  ... \\
\hline
\end{tabular}
\begin{center}
This table is available in its entirety in a machine-readable form
in the online journal. A portion is shown here for guidance 
regarding its form and content.
\end{center}
\end{table*}

\clearpage

\begin{figure*}
\centering
\includegraphics[angle=90,height=12cm]{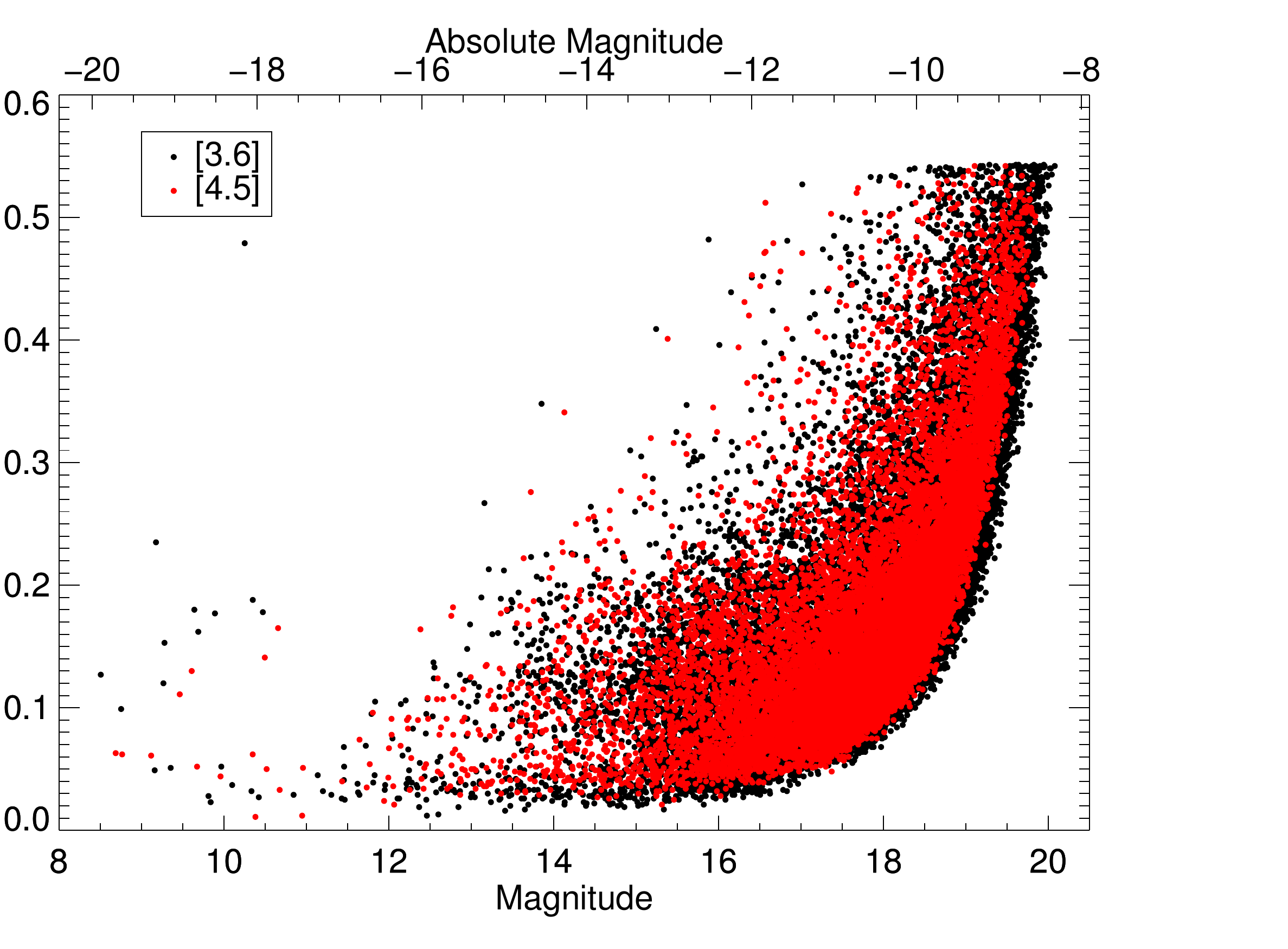}
\caption{Magnitude versus uncertainty for the \textit{Spitzer} [3.6]
  and [4.5] point sources. The [4.5] values are in red and show both
  the higher uncertainties and brighter limiting magnitude limit
  for this bandpass.}
\label{spitzuncerts}
\end{figure*}

\clearpage

\begin{figure*}
\begin{center}
{\includegraphics[angle=90,height=12cm]{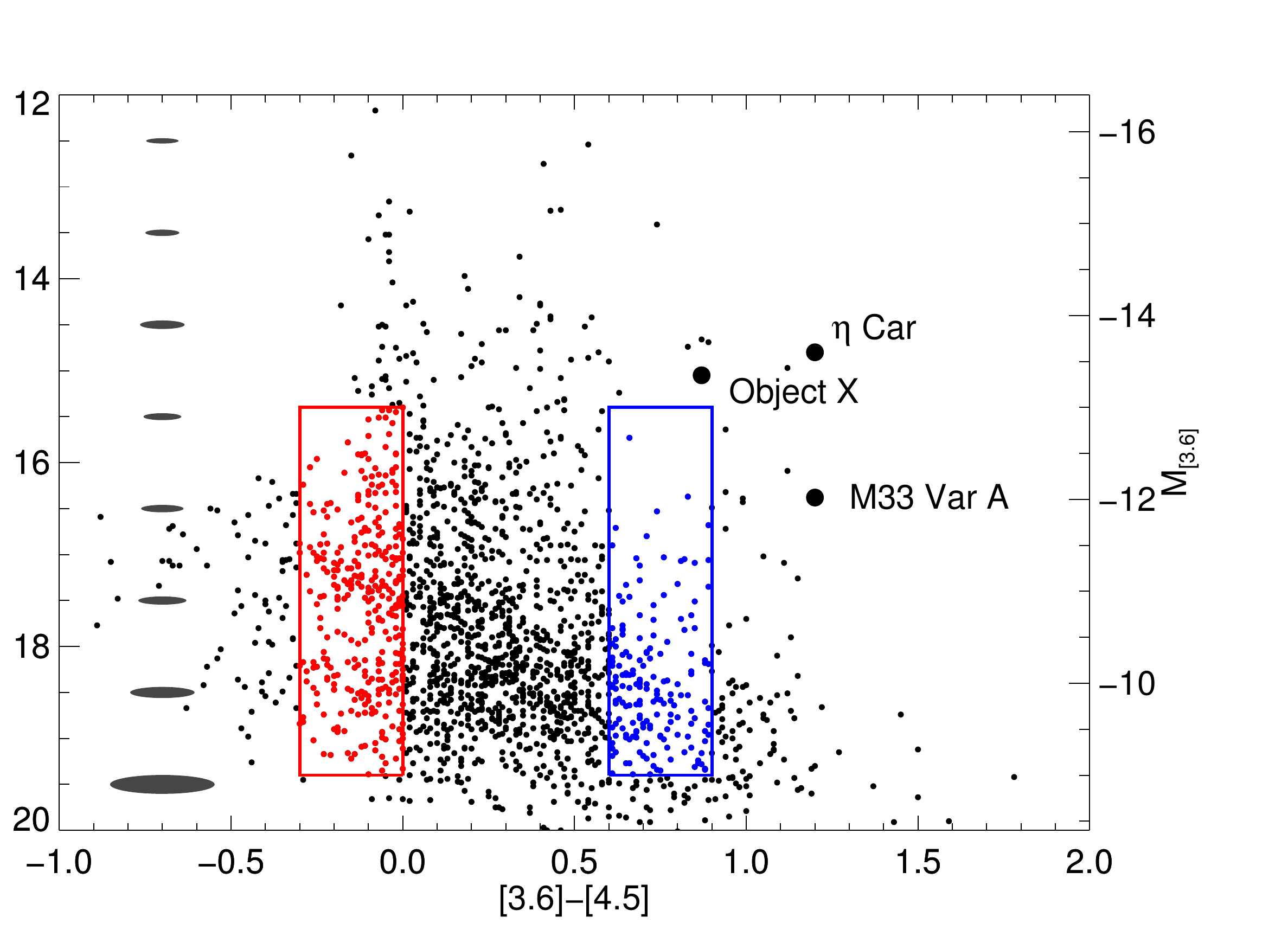}}
\end{center}
\caption{Color-magnitude diagram (CMD) for M83 based on \textit{Spitzer}
  IRAC Bands 1 and 2 photometry. Regions empirically known to contain 
  red supergiants (RSGs) and supergiant Be stars (sgB[e]) in 
  \citet{bon09,bon10} are outlined with red and blue boxes, 
  respectively. Objects that lie within those regions are plotted
  with red and blue points.
  On the left side of the plot are shown the 
  average 1$\sigma$ uncertainty ellipses for stars in 1 mag 
  bins for the [3.6] photometry and the [3.6]--[4.5] color. For
  reference, also plotted are the locations of the prominent 
  evolved massive stars $\eta$ Car \citep{hum94}, M33 Variable A
  \citep{hum06}, and Object X \citep{kha10} when placed at
  the distance ($\sim$4.8 Mpc) of M83.}
\label{3p6vcolor}
\end{figure*}

\clearpage

\begin{figure*}
\begin{center}
{\includegraphics[angle=90,height=12cm]{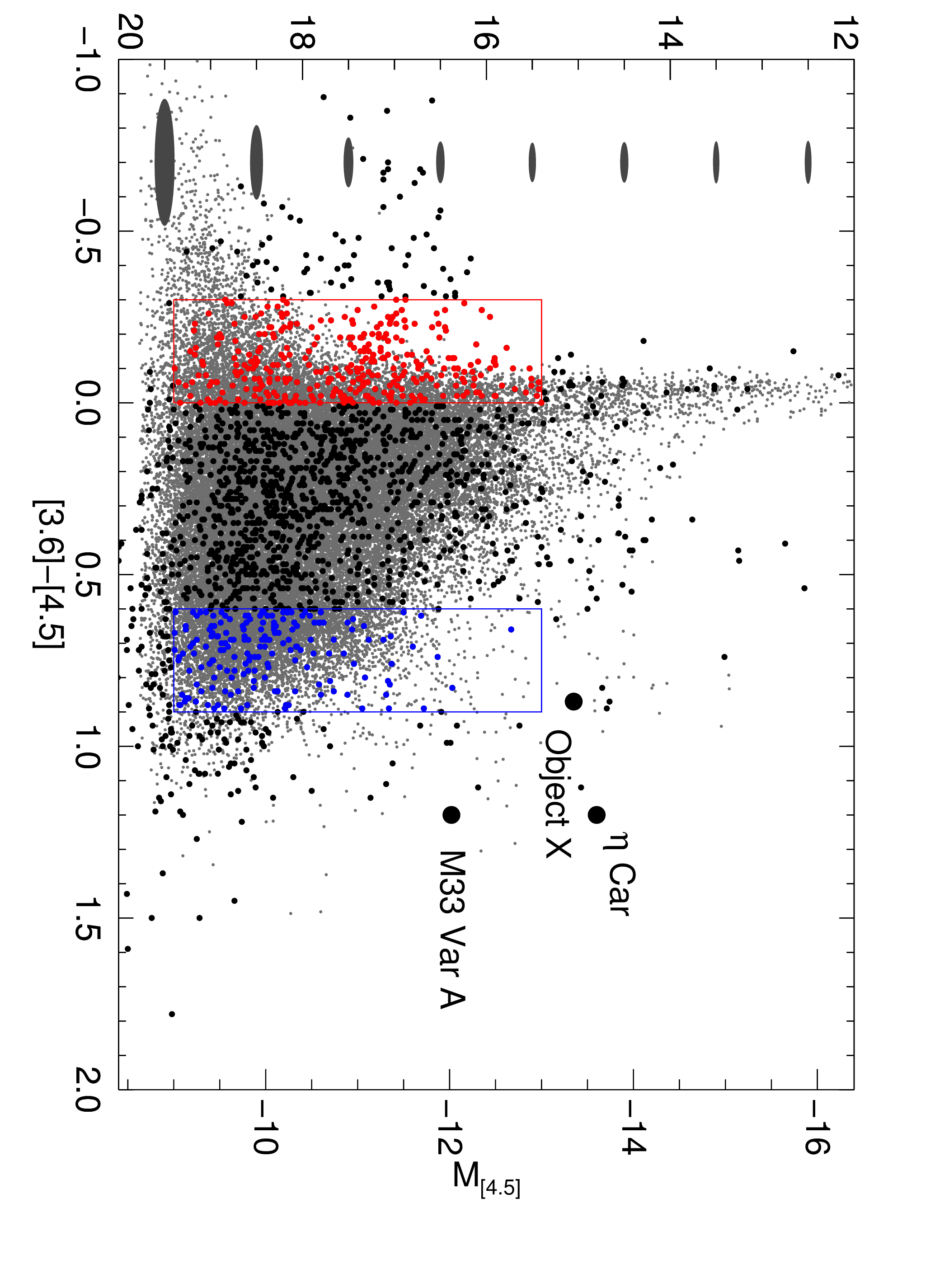}}
\end{center}
\caption{This figure is the same as Figure \ref{3p6vcolor}
  but includes grey points from the \textit{Spitzer} Deep Wide Field Survey
  representing the expected regions of contamination from non-M83 
  sources. Note that only the more luminous and bluer RSG 
  candidates lie distinctly outside the contaminated region. For
  reference, also plotted are the locations of the prominent 
  evolved massive stars $\eta$ Car \citep{hum94}, M33 Variable A
  \citep{hum06}, Object X \citep{kha11}, 
  when placed at the distance ($\sim$4.8 Mpc) of M83.}
\label{3p6vcolor-all}
\end{figure*}

\clearpage

\begin{figure*}
\begin{center}
{\includegraphics[angle=90,height=12cm]{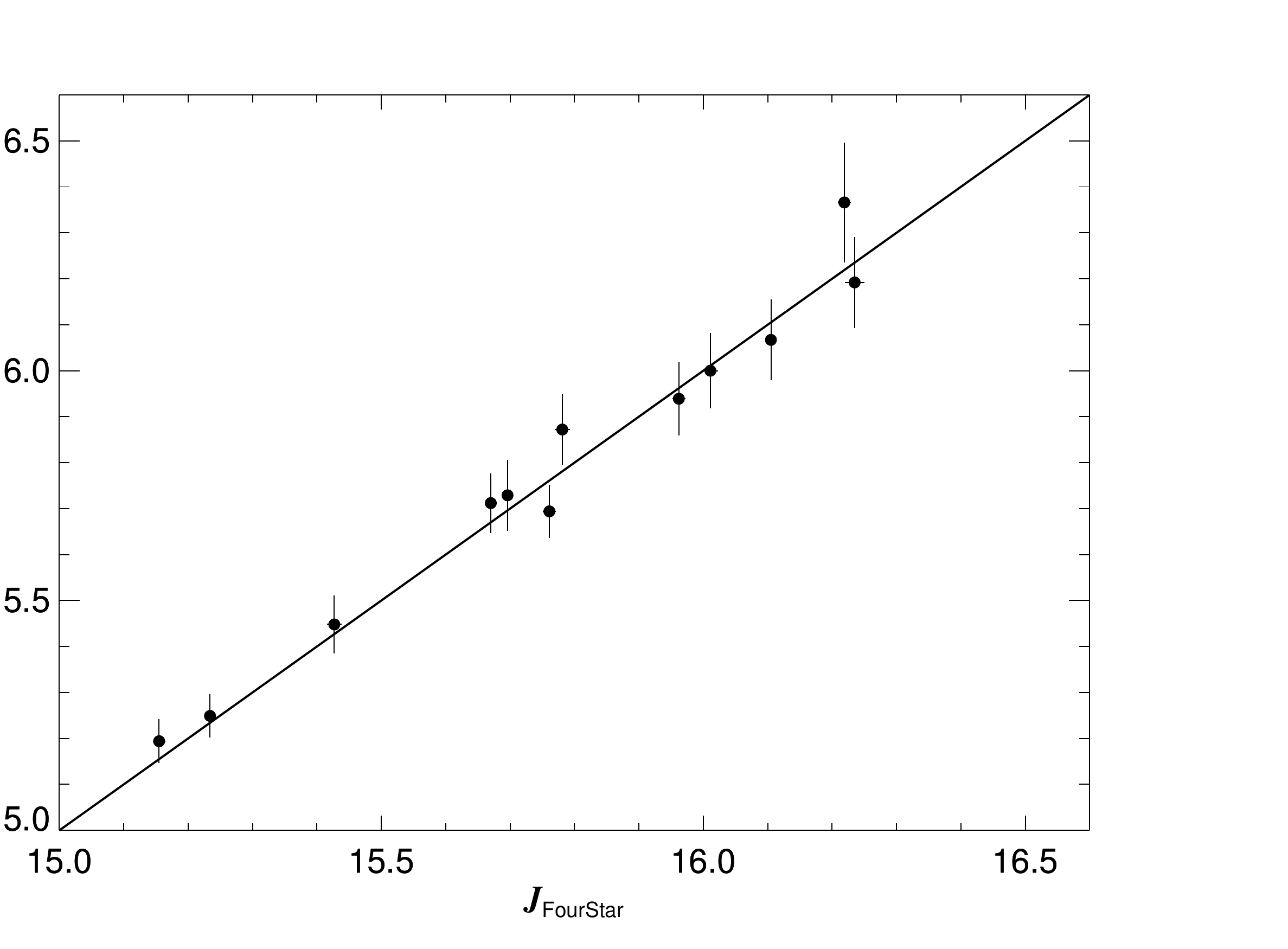}}
\end{center}
\caption{Comparison of the $J$ magnitude from FourStar PSF measurements to the
  $J$ magnitude from 2MASS. Note the much smaller uncertainties for 
  data from FourStar as compared to data from 2MASS.}
\label{jcompare}
\end{figure*}

\clearpage

\begin{figure*}
\begin{center}
{\includegraphics[angle=90,height=12cm]{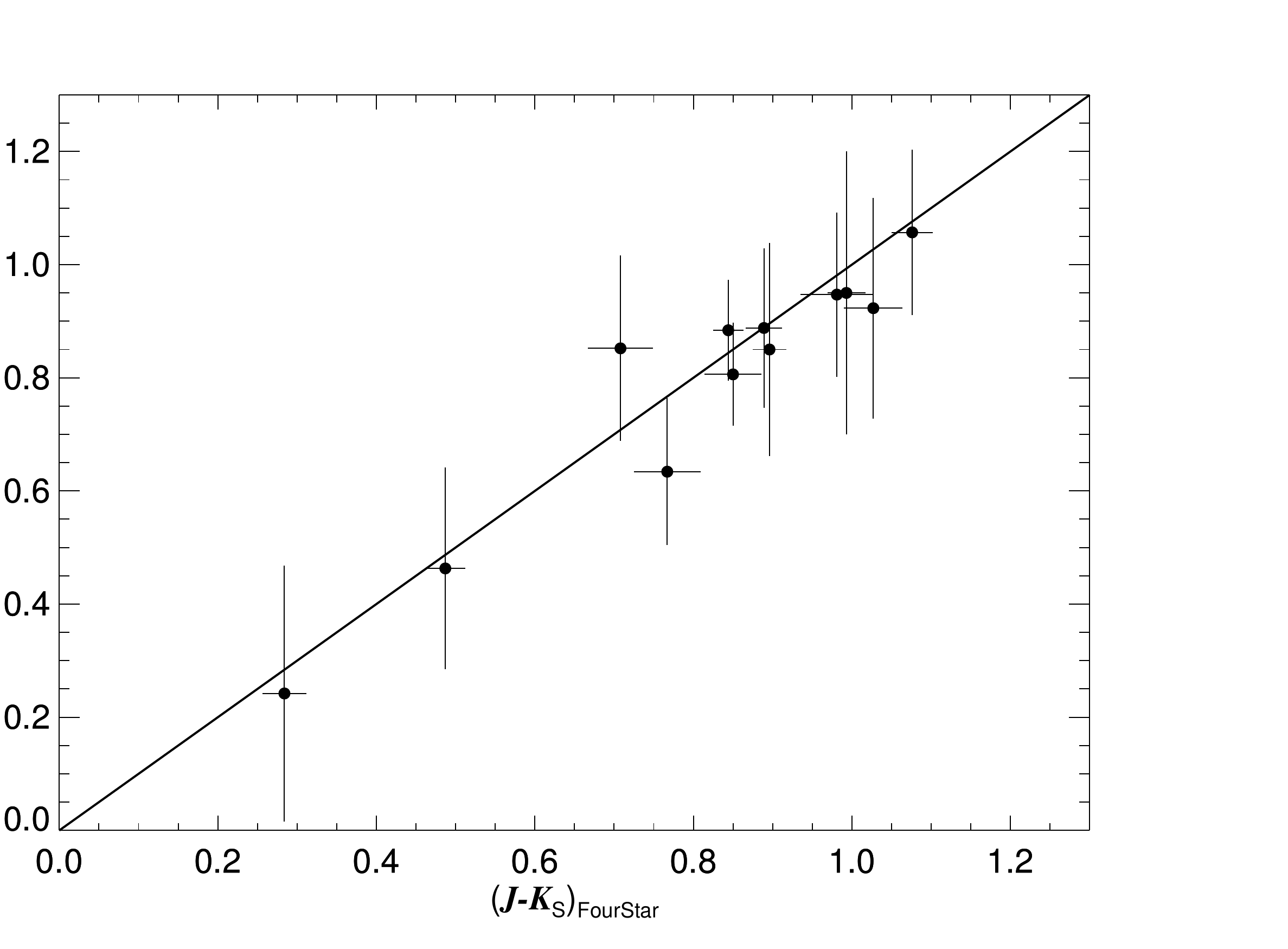}}
\end{center}
\caption{Comparison of the $(J-K_{\rm s})$ from FourStar PSF measurements 
  to the $(J-K_{\rm s})$ from 2MASS. Again, FourStar measurements have 
  much smaller uncertainties.}
\label{jmkcompare}
\end{figure*}

\clearpage

\begin{figure*}
\begin{center}
{\includegraphics[angle=90,height=12cm]{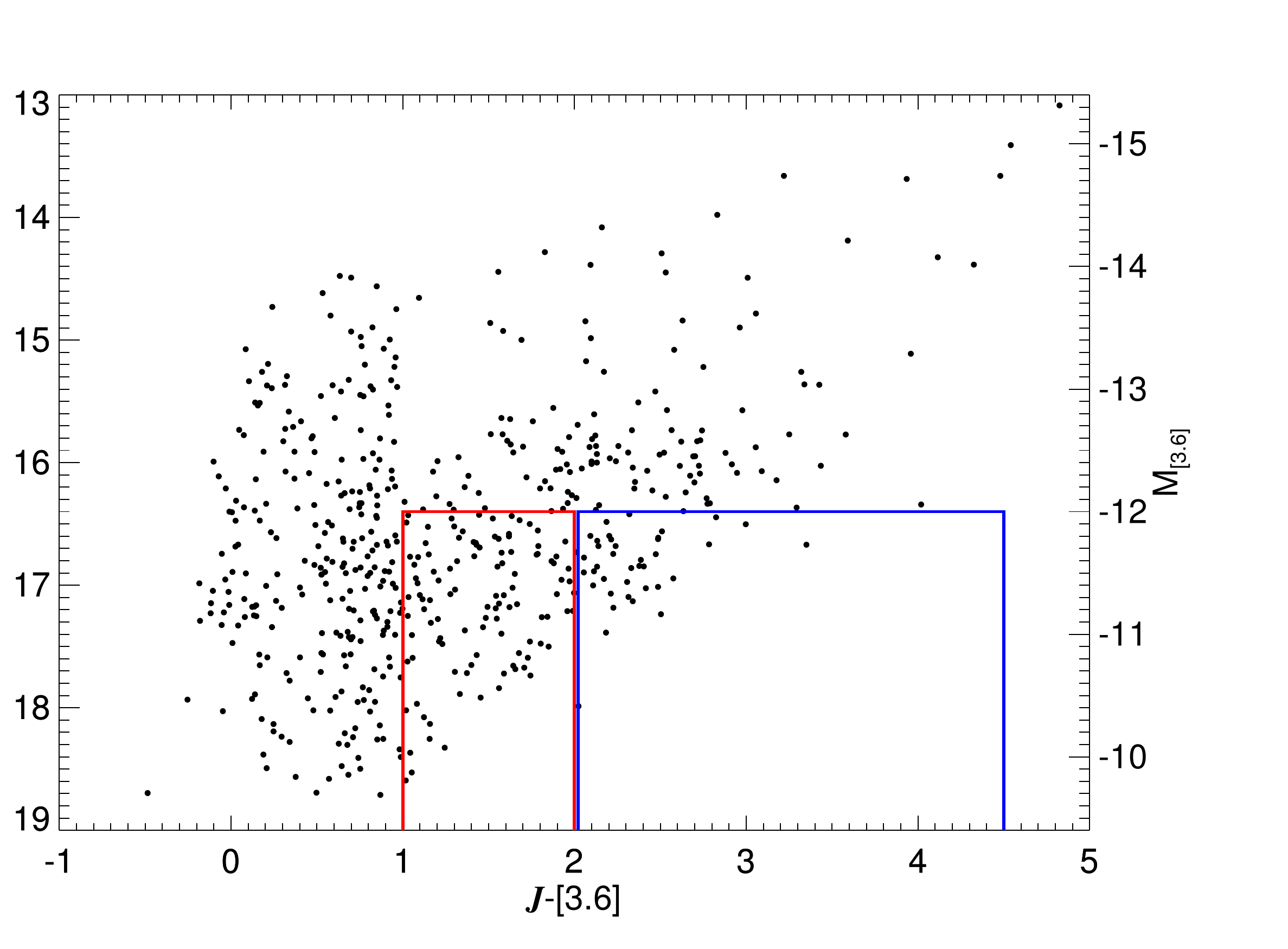}}
\end{center}
\caption{Color-magnitude diagram of [3.6] vs. $J-$[3.6] 
  for targets in the \textit{Spitzer} and near IR
  combined catalog. Regions of specific classes of stars as seen in
  Figure 3 of \citet{bon09,bon10} are 
  outlined by a red box for RSGs, and a blue box for sgB[e] stars.}
\label{jm3p6v3p6}
\end{figure*}

\clearpage

\begin{figure*}
\begin{center}
\subfloat[][Quality Rank 1]{\includegraphics[width=2in]{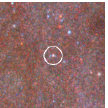}}
\subfloat[][Quality Rank 2]{\includegraphics[width=2in]{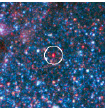}}\\
\subfloat[][Quality Rank 3]{\includegraphics[width=2in]{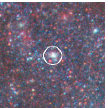}}
\subfloat[][Quality Rank 4]{\includegraphics[width=2in]{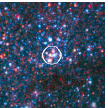}}\\
\subfloat[][Quality Rank 5]{\includegraphics[width=2in]{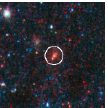}}
\end{center}
\caption{Composite \textit{HST} images (from F336W $U$, F438W $B$, F547M $y$,
  F555W $V$, F657N H$\alpha$, F814W $I$)
  of objects from Table \ref{maintable}
  demonstrating the ``Quality'' ranks listed in column 15.
  Both the F336W and F438W filter images make up the blue colors,
  F547M or F555W images are shown as green, while red colors are from
  F657N H$\alpha$ and F814W. 
  Each image is 10\arcsec~on a side. Over-plotted is the 1\farcs7
  \textit{Spitzer} point spread function. The quality ranks are: 
  (a) an isolated, excellent red candidate, rank 1, (b) an 
  extremely red candidate, rank 2, (c) an unresolved cluster or 
  other non-point source, rank 3, (d) a compact association, rank 4, 
  (e) a likely background galaxy, rank 5. In these images, North is up and
  East is left.}
\label{HSTimages}
\end{figure*}

\clearpage

\begin{figure*}
\begin{center}
{\includegraphics[]{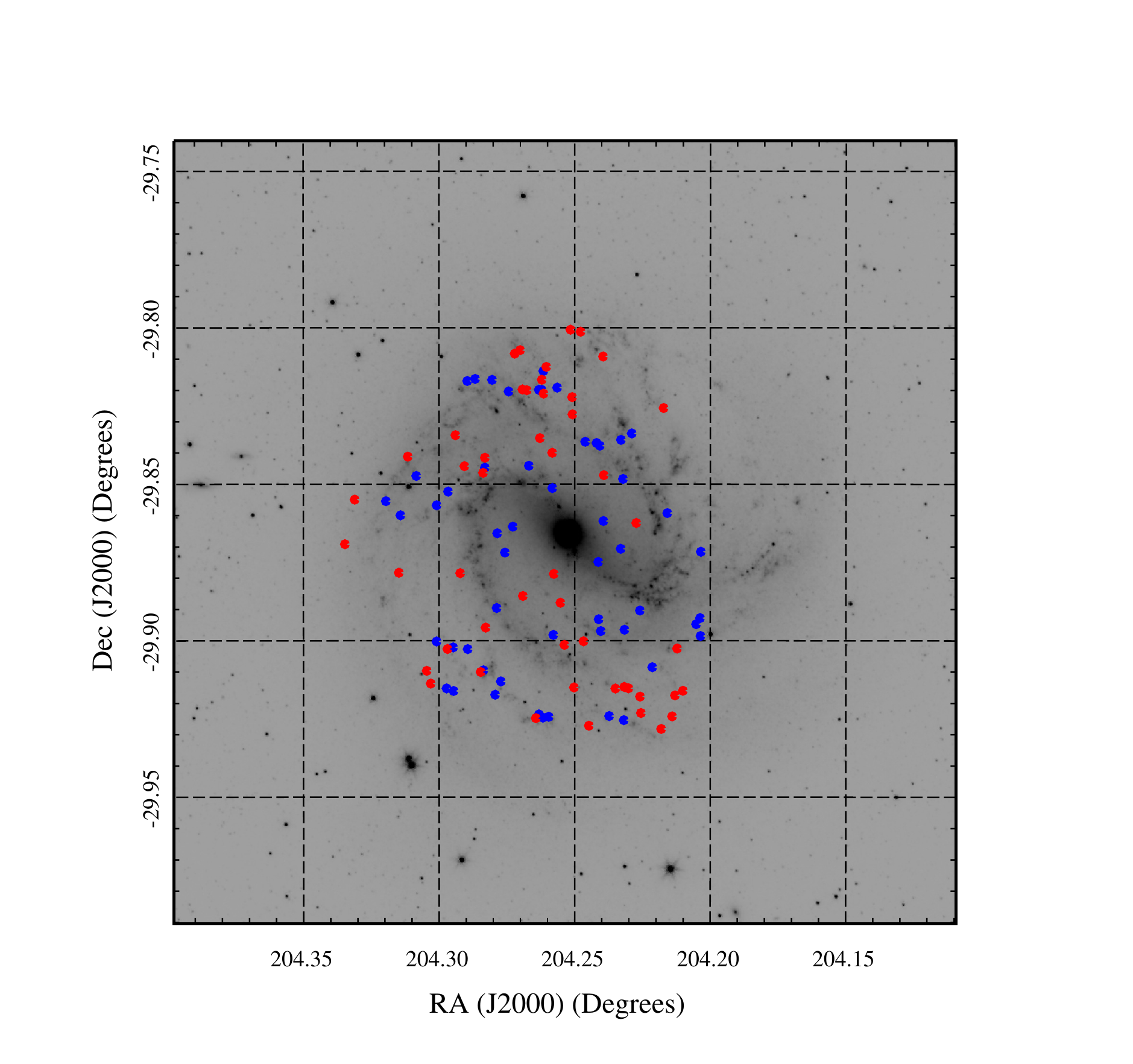}}
\end{center}
\caption{A 15$\arcmin$ $\times$ 15$\arcmin$ \textit{Spitzer} 3.6 $\mu$m 
  Band 1 image of M83, depicting objects with rank 1 and 2 as red circles,
  and rank 3 and higher shown as blue circles.
  All candidates lie within the \textit{HST} coverage and are in
  the spiral arms, where massive star candidates are most likely to reside.}
\label{m83pic}
\end{figure*}

\clearpage

\begin{figure*}
\begin{center}
{\includegraphics[angle=90,height=12cm]{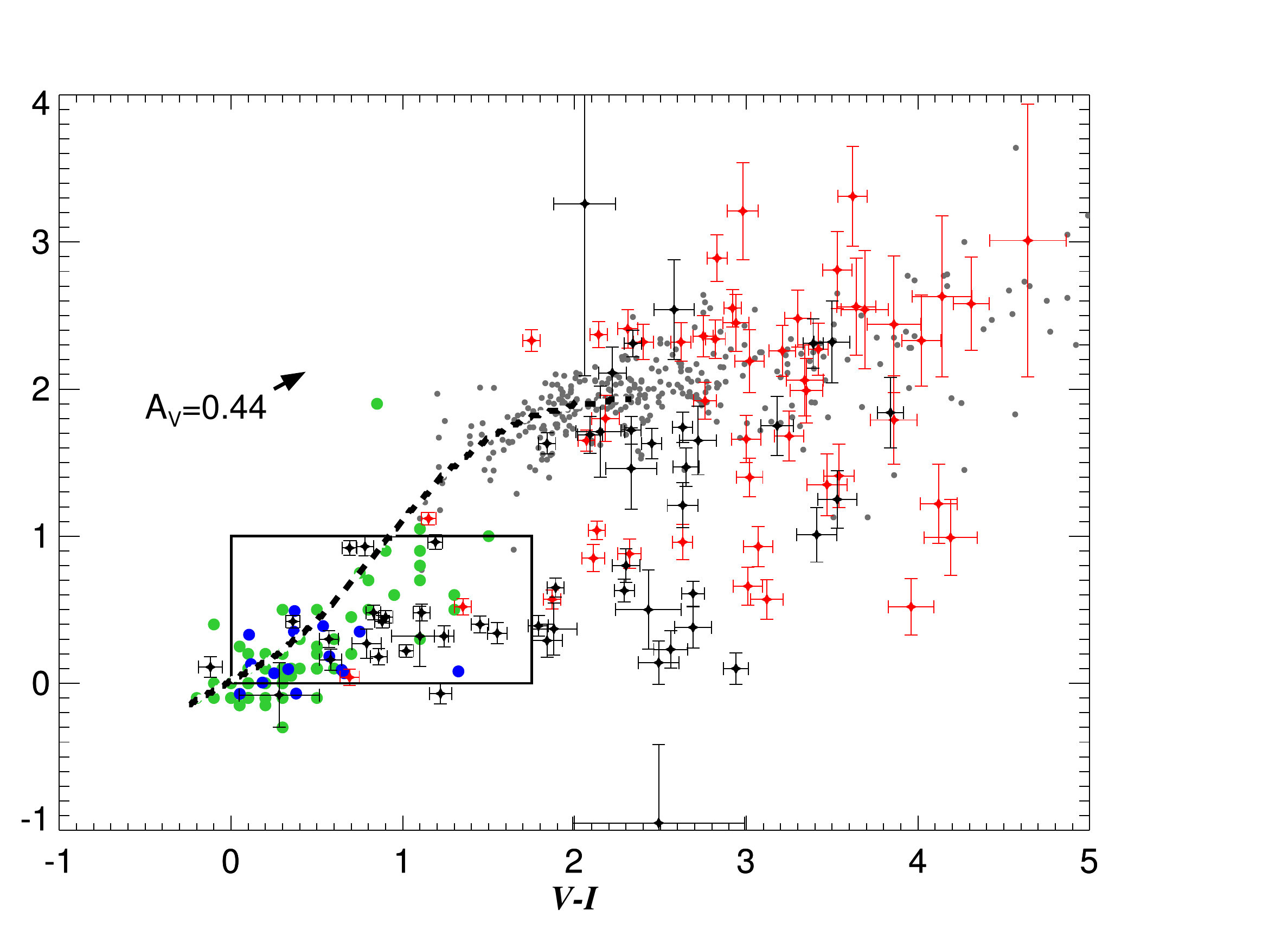}}
\end{center}
\caption{Color-color diagram for M83 stars with \textit{HST} photometry. 
  Objects
  determined to be good candidates for follow-up spectroscopy with rank
  1 and 2 are shown in red, while objects with ranks 3 and higher are
  shown in black.
  Also plotted is a ``typical'' $V$-band reddening for stars in
  M83's central region as reported by \citet{kim12}. Spectroscopically 
  confirmed RSGs 
  in the Local Group are collectively shown as grey points. The 
  dashed line represents the theoretical supergiant sequence from
  \citet{ber94} as used in \citet{gra13}. The box outlines the region
  where cluster candidates are known to exist \citep{bar06}. The blue
  points are sgB[e] stars from the LMC and SMC \citep{bon09,bon10}, 
  while the green points are blue supergiants from M31 and M33 
  \citep{hum14}.}
\label{bvvsvi}
\end{figure*}

\clearpage

\begin{figure*}
\begin{center}
{\includegraphics[angle=90,height=12cm]{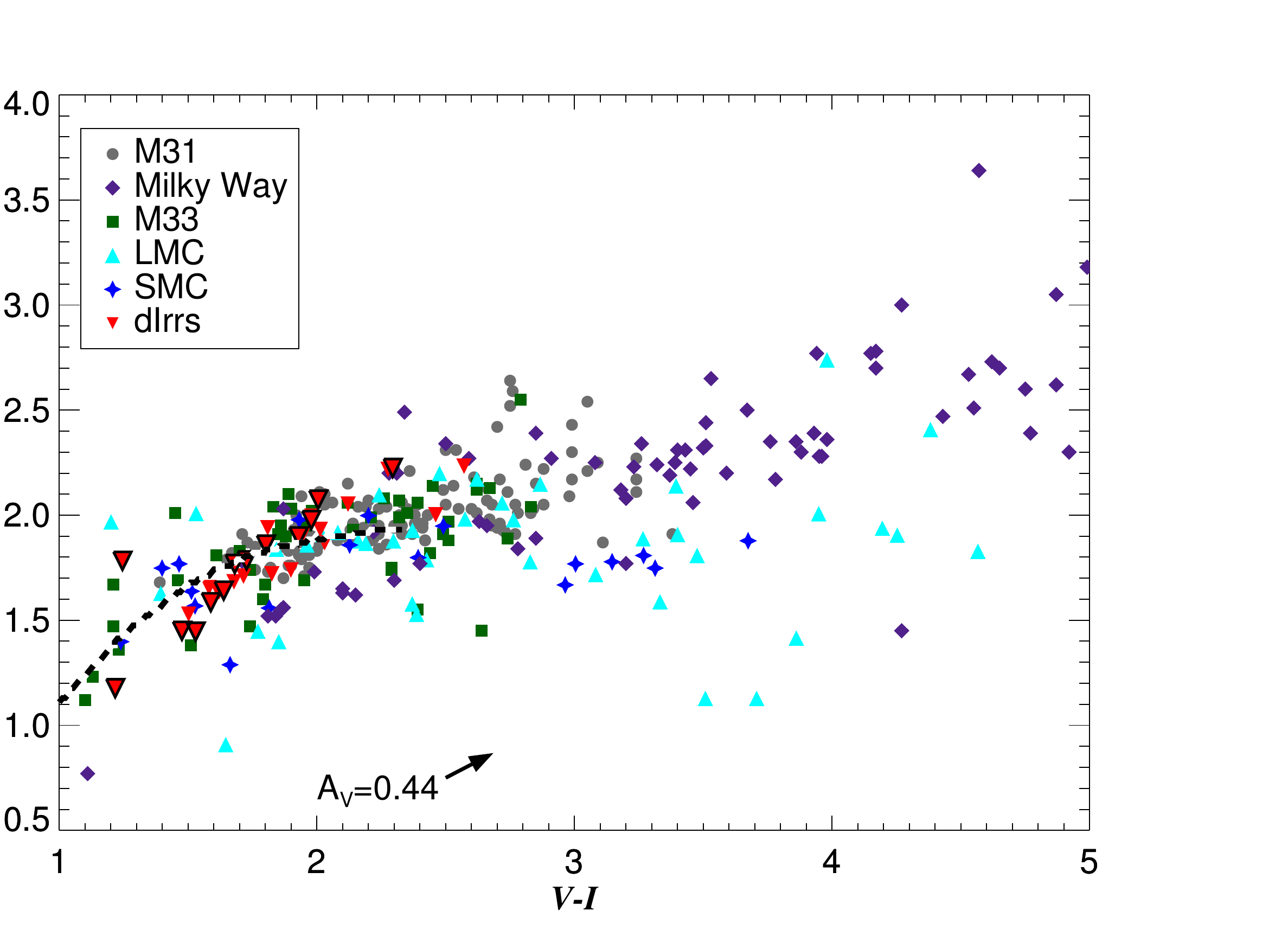}}
\end{center}
\caption{A color-color diagram showing the locations of spectroscopically
  confirmed RSGs broken down by galaxy in the Local Group. 
  Over-plotted is the theoretical
  supergiant sequence from \citet{gra13} and \citet{ber94}, as well as
  an arrow showing the typical reddening in M83 as determined by 
  \citet{kim12}.}
\label{bvvsvilocal}
\end{figure*}

\clearpage

\begin{figure*}
\begin{center}
{\includegraphics[angle=90,height=12cm]{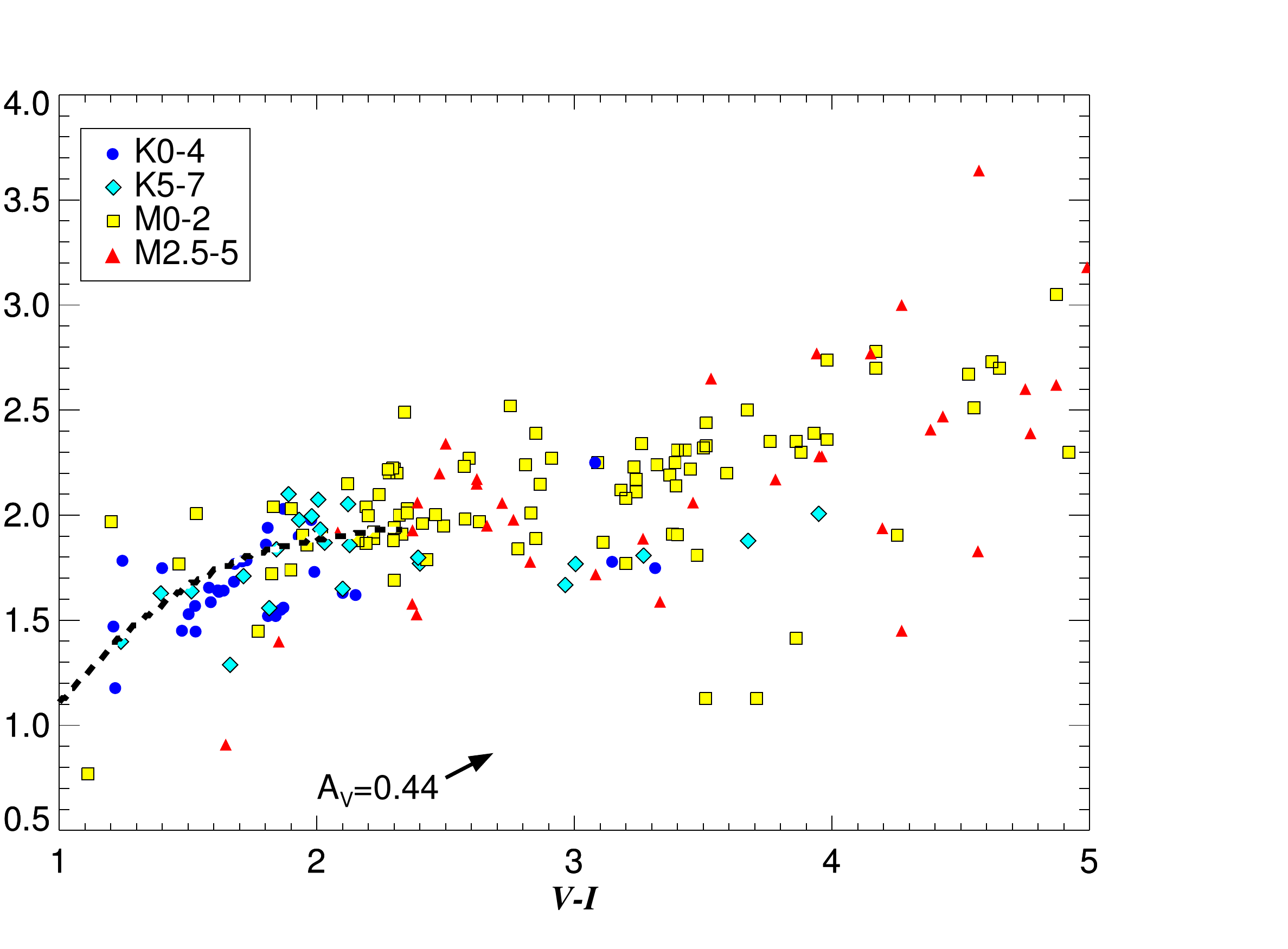}}
\end{center}
\caption{A color-color diagram showing the locations of spectroscopically
  confirmed RSGs in the Local Group broken down by spectral type.
  Over-plotted is the theoretical
  supergiant sequence from \citet{gra13} and \citet{ber94}.}
\label{bvvsvisptype}
\end{figure*}

\clearpage

\begin{figure*}
\begin{center}
\subfloat[][204.21721$-$29.82567]{\includegraphics[width=2in]{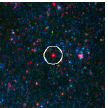}} 
\subfloat[][204.29392$-$29.83436]{\includegraphics[width=2in]{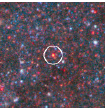}} 
\subfloat[][204.25825$-$29.83997]{\includegraphics[width=2in]{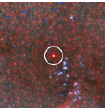}} \\
\subfloat[][204.31148$-$29.84119]{\includegraphics[width=2in]{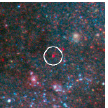}} 
\subfloat[][204.22728$-$29.86241]{\includegraphics[width=2in]{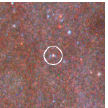}} 
\subfloat[][204.33467$-$29.86917]{\includegraphics[width=2in]{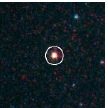}} \\
\subfloat[][204.33385$-$29.87635]{\includegraphics[width=2in]{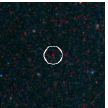}} 
\subfloat[][204.2828$-$29.89583]{\includegraphics[width=2in]{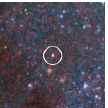}} 
\subfloat[][204.24671$-$29.90021]{\includegraphics[width=2in]{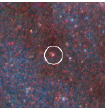}} \\
\subfloat[][204.30452$-$29.90964]{\includegraphics[width=2in]{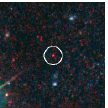}} 
\subfloat[][204.30305$-$29.91371]{\includegraphics[width=2in]{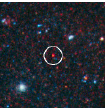}} 
\subfloat[][204.25023$-$29.91497]{\includegraphics[width=2in]{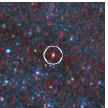}} 
\end{center}
\caption{Objects from Table \ref{maintable} with a rank of 1, thus 
  excellent candidates for a massive star. The color scheme for these 
  images is the same as for Figure \ref{HSTimages}.
  Each image is 10\arcsec~on a side, with the \textit{Spitzer} point 
  spread function of 1\farcs7 shown
  as a white circle. North is up and East is left.}
\end{figure*}

\clearpage


\begin{figure*}
\begin{center}
\subfloat[][204.21006$-$29.91601]{\includegraphics[width=2in]{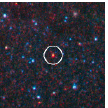}} 
\subfloat[][204.22586$-$29.91788]{\includegraphics[width=2in]{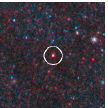}} 
\subfloat[][204.21403$-$29.9242]{\includegraphics[width=2in]{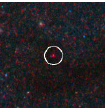}} \\
\subfloat[][204.2643$-$29.92477]{\includegraphics[width=2in]{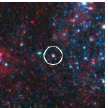}} 
\subfloat[][204.21807$-$29.92822]{\includegraphics[width=2in]{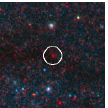}} 
\subfloat[][204.25152$-$29.80064]{\includegraphics[width=2in]{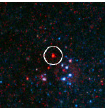}} \\
\subfloat[][204.23951$-$29.80921]{\includegraphics[width=2in]{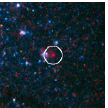}} 
\subfloat[][204.26044$-$29.81253]{\includegraphics[width=2in]{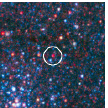}} 
\subfloat[][204.26763$-$29.81998]{\includegraphics[width=2in]{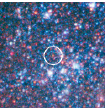}} \\
\subfloat[][204.26145$-$29.82107]{\includegraphics[width=2in]{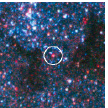}} 
\subfloat[][204.25092$-$29.82219]{\includegraphics[width=2in]{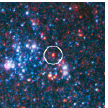}} 
\subfloat[][204.2508$-$29.8277]{\includegraphics[width=2in]{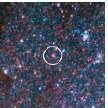}} \\
\end{center}
\caption{More objects from Table \ref{maintable} with a rank of 1, thus excellent candidates for a massive star.
  The color scheme for these images is the same as for Figure \ref{HSTimages}.
  Each image is 10\arcsec~on a side, with the \textit{Spitzer} point spread function of 1\farcs7 shown
  as a white circle. North is up and East is left.}
\end{figure*}

\clearpage

\begin{figure*}
\begin{center}
\subfloat[][204.2628$-$29.83529]{\includegraphics[width=2in]{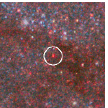}} 
\subfloat[][204.28305$-$29.84155]{\includegraphics[width=2in]{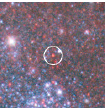}} 
\subfloat[][204.29062$-$29.84425]{\includegraphics[width=2in]{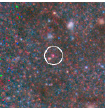}} \\
\subfloat[][204.28372$-$29.84638]{\includegraphics[width=2in]{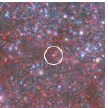}} 
\subfloat[][204.33106$-$29.85492]{\includegraphics[width=2in]{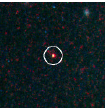}} 
\subfloat[][204.29217$-$29.87845]{\includegraphics[width=2in]{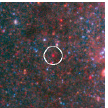}} \\
\subfloat[][204.25766$-$29.87868]{\includegraphics[width=2in]{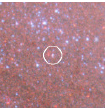}} 
\subfloat[][204.26906$-$29.88573]{\includegraphics[width=2in]{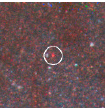}} 
\subfloat[][204.25522$-$29.88789]{\includegraphics[width=2in]{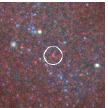}} \\
\subfloat[][204.21221$-$29.90251]{\includegraphics[width=2in]{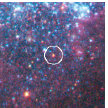}} 
\subfloat[][204.27102$-$29.90557]{\includegraphics[width=2in]{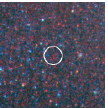}} 
\subfloat[][204.28463$-$29.91002]{\includegraphics[width=2in]{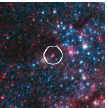}} \\
\end{center}
\caption{More objects from Table \ref{maintable} with a rank of 1, thus excellent candidates for a massive star.
  The color scheme for these images is the same as for Figure \ref{HSTimages}.
  Each image is 10\arcsec~on a side, with the \textit{Spitzer} point spread function of 1\farcs7 shown
  as a white circle. North is up and East is left.}
\end{figure*}

\clearpage

\begin{figure*}
\begin{center}
\subfloat[][204.2317$-$29.9148]{\includegraphics[width=2in]{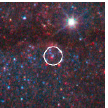}} 
\subfloat[][204.21299$-$29.91749]{\includegraphics[width=2in]{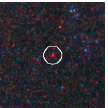}} 
\subfloat[][204.22553$-$29.92313]{\includegraphics[width=2in]{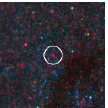}} \\
\subfloat[][204.24474$-$29.92718]{\includegraphics[width=2in]{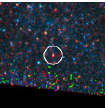}} \\
\end{center}
\caption{More objects from Table \ref{maintable} with a rank of 1, thus excellent candidates for a massive star.
  The color scheme for these images is the same as for Figure \ref{HSTimages}.
  Each image is 10\arcsec~on a side, with the \textit{Spitzer} point spread function of 1\farcs7 shown
  as a white circle. North is up and East is left.}
\end{figure*}



\begin{figure*}
\begin{center}
\subfloat[][204.24779$-$29.80129]{\includegraphics[width=2in]{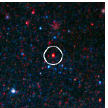}} 
\subfloat[][204.27008$-$29.80718]{\includegraphics[width=2in]{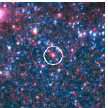}} 
\subfloat[][204.27208$-$29.80829]{\includegraphics[width=2in]{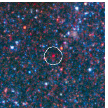}} \\
\subfloat[][204.26209$-$29.81663]{\includegraphics[width=2in]{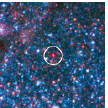}} 
\subfloat[][204.26922$-$29.81975]{\includegraphics[width=2in]{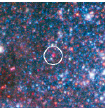}} 
\subfloat[][204.23924$-$29.84713]{\includegraphics[width=2in]{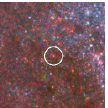}} \\
\subfloat[][204.21703$-$29.87143]{\includegraphics[width=2in]{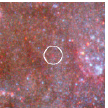}} 
\subfloat[][204.31479$-$29.8783]{\includegraphics[width=2in]{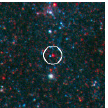}} 
\subfloat[][204.26852$-$29.89639]{\includegraphics[width=2in]{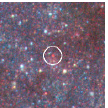}} \\
\subfloat[][204.25377$-$29.90135]{\includegraphics[width=2in]{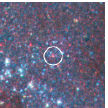}} 
\subfloat[][204.22208$-$29.90175]{\includegraphics[width=2in]{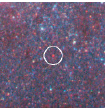}} 
\subfloat[][204.29688$-$29.90267]{\includegraphics[width=2in]{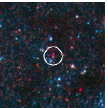}} \\
\end{center}
\caption{Objects from Table \ref{maintable} with a rank of 2, thus decent candidates for a massive star.
  The color scheme for these images is the same as for Figure \ref{HSTimages}.
  Each image is 10\arcsec~on a side, with the \textit{Spitzer} point spread function of 1\farcs7 shown
  as a white circle. North is up and East is left.}
\end{figure*}

\clearpage

\begin{figure*}
\begin{center}
\subfloat[][204.23015$-$29.91522]{\includegraphics[width=2in]{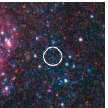}} 
\subfloat[][204.23495$-$29.9153]{\includegraphics[width=2in]{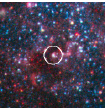}} 
\subfloat[][204.21763$-$29.92694]{\includegraphics[width=2in]{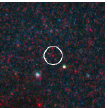}} \\
\subfloat[][204.28307$-$29.84473]{\includegraphics[width=2in]{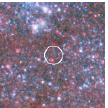}} 
\subfloat[][204.30836$-$29.84737]{\includegraphics[width=2in]{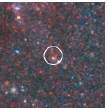}} 
\subfloat[][204.25821$-$29.85125]{\includegraphics[width=2in]{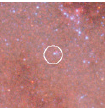}} \\
\subfloat[][204.31955$-$29.85543]{\includegraphics[width=2in]{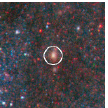}} 
\subfloat[][204.31419$-$29.85995]{\includegraphics[width=2in]{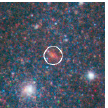}} 
\subfloat[][204.23301$-$29.87065]{\includegraphics[width=2in]{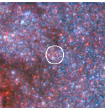}} \\
\subfloat[][204.20346$-$29.87159]{\includegraphics[width=2in]{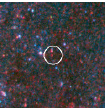}} \\
\end{center}
\caption{More objects from Table \ref{maintable} with a rank of 2, thus decent candidates for a massive star.
  The color scheme for these images is the same as for Figure \ref{HSTimages}.
  Each image is 10\arcsec~on a side, with the \textit{Spitzer} point spread function of 1\farcs7 shown
  as a white circle. North is up and East is left.}
\end{figure*}

\clearpage

\begin{figure*}
\begin{center}
\subfloat[][204.28665$-$29.81637]{\includegraphics[width=2in]{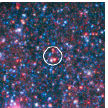}} 
\subfloat[][204.28041$-$29.81667]{\includegraphics[width=2in]{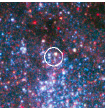}} 
\subfloat[][204.28956$-$29.81703]{\includegraphics[width=2in]{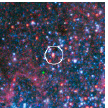}} \\
\subfloat[][204.25643$-$29.81915]{\includegraphics[width=2in]{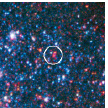}} 
\subfloat[][204.26219$-$29.81984]{\includegraphics[width=2in]{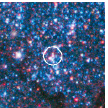}} 
\subfloat[][204.26318$-$29.81985]{\includegraphics[width=2in]{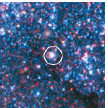}} \\
\subfloat[][204.27432$-$29.82038]{\includegraphics[width=2in]{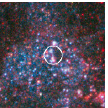}} 
\subfloat[][204.24071$-$29.83772]{\includegraphics[width=2in]{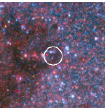}} 
\subfloat[][204.26684$-$29.84411]{\includegraphics[width=2in]{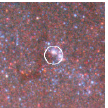}} \\
\subfloat[][204.29662$-$29.85237]{\includegraphics[width=2in]{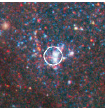}} 
\subfloat[][204.23938$-$29.86178]{\includegraphics[width=2in]{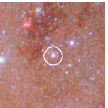}} 
\subfloat[][204.27568$-$29.87186]{\includegraphics[width=2in]{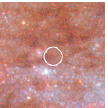}} \\
\end{center}
\caption{Objects from Table \ref{maintable} with a rank of 3, thus not good candidates for a massive star.
  The color scheme for these images is the same as for Figure \ref{HSTimages}.
  Each image is 10\arcsec~on a side, with the \textit{Spitzer} point spread function of 1\farcs7 shown
  as a white circle. North is up and East is left.}
\end{figure*}

\clearpage

\begin{figure*}
\begin{center}
\subfloat[][204.30318$-$29.87883]{\includegraphics[width=2in]{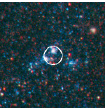}} 
\subfloat[][204.27873$-$29.88957]{\includegraphics[width=2in]{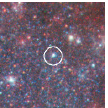}} 
\subfloat[][204.22592$-$29.89037]{\includegraphics[width=2in]{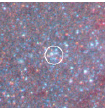}} \\
\subfloat[][204.24112$-$29.89317]{\includegraphics[width=2in]{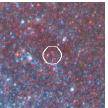}} 
\subfloat[][204.24035$-$29.89695]{\includegraphics[width=2in]{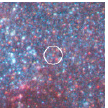}} 
\subfloat[][204.25784$-$29.89818]{\includegraphics[width=2in]{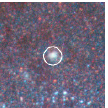}} \\
\subfloat[][204.20358$-$29.89852]{\includegraphics[width=2in]{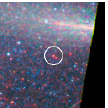}} 
\subfloat[][204.28941$-$29.9027]{\includegraphics[width=2in]{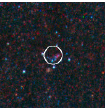}} 
\subfloat[][204.27719$-$29.91304]{\includegraphics[width=2in]{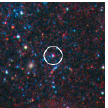}} \\
\subfloat[][204.29721$-$29.91526]{\includegraphics[width=2in]{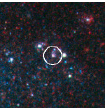}} 
\subfloat[][204.29463$-$29.91608]{\includegraphics[width=2in]{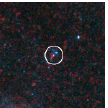}} 
\subfloat[][204.27929$-$29.91728]{\includegraphics[width=2in]{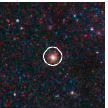}} \\
\end{center}
\caption{More objects from Table \ref{maintable} with a rank of 3, thus not good candidates for a massive star.
  The color scheme for these images is the same as for Figure \ref{HSTimages}.
  Each image is 10\arcsec~on a side, with the \textit{Spitzer} point spread function of 1\farcs7 shown
  as a white circle. North is up and East is left.}
\end{figure*}

\clearpage

\begin{figure*}
\begin{center}
\subfloat[][204.26318$-$29.92358]{\includegraphics[width=2in]{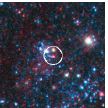}} 
\subfloat[][204.26171$-$29.9246]{\includegraphics[width=2in]{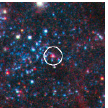}} 
\subfloat[][204.24593$-$29.92507]{\includegraphics[width=2in]{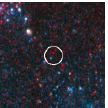}} \\
\subfloat[][204.23161$-$29.89652]{\includegraphics[width=2in]{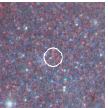}} \\
\end{center}
\caption{More objects from Table \ref{maintable} with a rank of 3, thus not good candidates for a massive star.
  The color scheme for these images is the same as for Figure \ref{HSTimages}.
  Each image is 10\arcsec~on a side, with the \textit{Spitzer} point spread function of 1\farcs7 shown
  as a white circle. North is up and East is left.}
\end{figure*}

\clearpage

\begin{figure*}
\begin{center}
\subfloat[][204.26152$-$29.81385]{\includegraphics[width=2in]{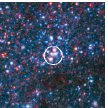}} 
\subfloat[][204.22893$-$29.8338]{\includegraphics[width=2in]{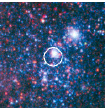}} 
\subfloat[][204.23292$-$29.83584]{\includegraphics[width=2in]{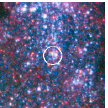}} \\
\subfloat[][204.24608$-$29.83645]{\includegraphics[width=2in]{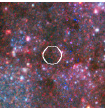}} 
\subfloat[][204.24191$-$29.83685]{\includegraphics[width=2in]{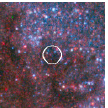}} 
\subfloat[][204.30085$-$29.85674]{\includegraphics[width=2in]{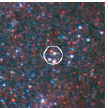}} \\
\subfloat[][204.24128$-$29.87485]{\includegraphics[width=2in]{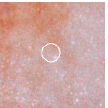}} 
\subfloat[][204.20519$-$29.89473]{\includegraphics[width=2in]{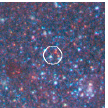}} 
\subfloat[][204.3009$-$29.90026]{\includegraphics[width=2in]{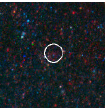}} \\
\subfloat[][204.28363$-$29.90948]{\includegraphics[width=2in]{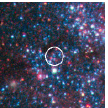}} 
\subfloat[][204.23725$-$29.9241]{\includegraphics[width=2in]{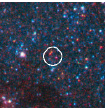}} 
\subfloat[][204.23186$-$29.92541]{\includegraphics[width=2in]{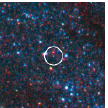}} \\
\end{center}
\caption{Objects from Table \ref{maintable} with a rank of 4, thus very poor candidates for a massive star.
  The color scheme for these images is the same as for Figure \ref{HSTimages}.
  Each image is 10\arcsec~on a side, with the \textit{Spitzer} point spread function of 1\farcs7 shown
  as a white circle. North is up and East is left.}
\end{figure*}

\clearpage

\begin{figure*}
\begin{center}
\subfloat[][204.2652$-$29.81649]{\includegraphics[width=2in]{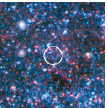}} 
\subfloat[][204.23216$-$29.84833]{\includegraphics[width=2in]{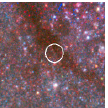}} 
\subfloat[][204.21583$-$29.85924]{\includegraphics[width=2in]{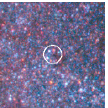}} \\
\subfloat[][204.27851$-$29.8657]{\includegraphics[width=2in]{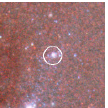}} 
\subfloat[][204.20375$-$29.89278]{\includegraphics[width=2in]{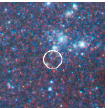}} 
\subfloat[][204.25955$-$29.92432]{\includegraphics[width=2in]{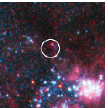}} \\
\end{center}
\caption{More objects from Table \ref{maintable} with a rank of 4, thus very poor candidates for a massive star.
  The color scheme for these images is the same as for Figure \ref{HSTimages}.
  Each image is 10\arcsec~on a side, with the \textit{Spitzer} point spread function of 1\farcs7 shown
  as a white circle. North is up and East is left.}
\end{figure*}

\clearpage

\begin{figure*}
\begin{center}
\subfloat[][204.25765$-$29.81874]{\includegraphics[width=2in]{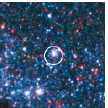}} 
\subfloat[][204.27278$-$29.86357]{\includegraphics[width=2in]{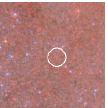}} 
\subfloat[][204.29479$-$29.90216]{\includegraphics[width=2in]{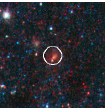}} \\
\subfloat[][204.22135$-$29.90849]{\includegraphics[width=2in]{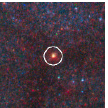}} \\
\end{center}
\caption{Objects from Table \ref{maintable} with a rank of 5, thus extremely poor candidates for a massive star.
  The color scheme for these images is the same as for Figure \ref{HSTimages}.
  Each image is 10\arcsec~on a side, with the \textit{Spitzer} point spread function of 1\farcs7 shown
  as a white circle. North is up and East is left.}
\end{figure*}

\end{document}